\documentclass[twocolumn,tighten]{aastex63}

\newcommand{\beam}{$\theta_{\mbox{\scriptsize maj}}\times\theta_{\mbox{\scriptsize min}}$}

\newcommand{\ujyperbeam}{$\mu$Jy\,beam$^{-1}$}

\usepackage{threeparttable}
\usepackage{wrapfig}
\usepackage{comment}

\accepted{June 15, 2023}
\submitjournal{ApJ}

\shorttitle{Anisotropic Ionizing Illumination from an M-type Pre-main Sequence Star}
\shortauthors{Terada et al.}
\graphicspath{{./}{figures/}}
\begin{document}

\title{Anisotropic Ionizing Illumination from an M-type Pre-main Sequence Star, DM Tau}

\correspondingauthor{Hauyu Baobab Liu}
\email{baobabyoo@gmail.com}

\author[0000-0003-2887-6381]{Yuka Terada}
%\affil{Physics Department, National Sun Yat-Sen University, No. 70, Lien-Hai Road, Kaosiung City 80424, Taiwan, R.O.C.}
\affil{Institute of Astronomy and Astrophysics, Academia Sinica, 11F of Astronomy-Mathematics Building, AS/NTU No.1, Sec. 4,
Roosevelt Rd, Taipei 10617, Taiwan, ROC}
\affil{Department of Astrophysics, National Taiwan University, Taipei 10617, Taiwan, R.O.C.} 

%\collaboration{1}{(AAS Journals Data Scientists collaboration)}

\author[0000-0003-2300-2626]{Hauyu Baobab Liu}
 \affil{Department of Physics, National Sun Yat-Sen University, No. 70, Lien-Hai Road, Kaohsiung City 80424, Taiwan, R.O.C.}
% \affil{Institute of Astronomy and Astrophysics, Academia Sinica, 11F of Astronomy-Mathematics Building, AS/NTU No.1, Sec. 4, Roosevelt Rd, Taipei 10617, Taiwan, ROC} 

\author{David Mkrtichian}
\affil{National Astronomical Research Institute of Thailand, 260  Moo 4, T. Donkaew,  A. Maerim, Chiangmai, 50180, Thailand}

\author[0000-0003-4361-5577]{Jinshi Sai}
\affil{Institute of Astronomy and Astrophysics, Academia Sinica, 11F of Astronomy-Mathematics Building, AS/NTU No.1, Sec. 4,
Roosevelt Rd, Taipei 10617, Taiwan, ROC} 

\author[0000-0003-0114-0542]{Mihoko Konishi}
\affil{Faculty of Science and Technology, Oita University, 700 Dannoharu, Oita, 870-1192, Japan}

% \author{Kai-Yang Lin}https://www.overleaf.com/project/5e1d61c8b493c600012bf9f1
% \affil{Institute of Astronomy and Astrophysics, Academia Sinica, 11F of Astronomy-Mathematics Building, AS/NTU No.1, Sec. 4,
% Roosevelt Rd, Taipei 10617, Taiwan, ROC} 

\author[0000-0001-7359-3300]{Ing-Guey Jiang}
\affil{Department of Physics and Institute of Astronomy, National Tsing Hua University, Hsinchu 30013, Taiwan}

\author{Takayuki Muto}
\affil{Kogakuin University, 2665-1 Nakano, Hachioji, Tokyo 192-0015, Japan}

\author[0000-0002-3053-3575]{Jun Hashimoto}
\affil{Astrobiology Center, 2-21-1 Osawa, Mitaka, Tokyo 181-8588, Japan}
\affil{Subaru Telescope, National Astronomical Observatory of Japan, Mitaka, Tokyo, 181-8588, Japan}
\affil{Department of Astronomy, School of Science, Graduate University for Advanced Studies (SOKENDAI), Mitaka, Tokyo 181-8588, Japan}

\author[0000-0002-6510-0681]{Motohide Tamura}
\affil{Astrobiology Center, 2-21-1 Osawa, Mitaka, Tokyo 181-8588, Japan}
\affil{Department of Astronomy, Graduate School of Science, The University of Tokyo, Hongo, Bunkyo-ku, Tokyo, Japan}
\affil{National Astronomical Observatory of Japan, Osawa, Mitaka Tokyo, Japan}

%\nocollaboration{0}



\begin{abstract}
The powerful, high-energy magnetic activities of young stars play important roles in the magnetohydrodynamics in the innermost parts of the protoplanetary disks. 
In addition, the associated UV and X-ray emission dictates the photochemistry; moreover, the corona activities can affect the atmosphere of a newborn extra-solar planet. 
How the UV and X-ray photons are generated, and how they illuminate the disks, are not well understood. 
Here we report the analyses of the optical and infrared (OIR) photometric monitoring observations and the high angular-resolution centimeter band images of the low-mass (M1 type) pre-main sequence star, DM Tau. 
We found that the OIR photometric light curves present periodic variations, which is consistent with that the host young star is rotating in the same direction as the natal disk and is hosting at least one giant cold spot. 
In addition, we resolved that the ionized gas in the DM Tau disk is localized, and its spatial distribution is varying with time. 
All the present observations can be coherently interpreted, if the giant cold spot is the dominant anisotropic UV and/or X-ray source that illuminates the ambient cone-like region. 
These results indicate that a detailed theoretical model of the high-energy protostellar emission is essential in the understanding of the space weather around the extra-solar planets and the origin of life.
\end{abstract}

\keywords{Protoplanetary disks(1300); Star-planet interactions(2177); Starspots(1572); Stellar accretion disks(1579); T Tauri stars(1681); Observational astronomy(1145) 
%\textcolor{red}{The rule for keywords has changed. Please follow the instruction here https://journals.aas.org/news/aas-journals-uat/ }
}


\section{Introduction} \label{sec:intro}

\begin{deluxetable*}{ l l c c c c }
\tablecaption{JVLA observations on DM~Tau \label{tab:JVLAobs}}
\tabletypesize{\scriptsize}
\tablecolumns{6}
%\tablenum{1}
%\tablewidth{10pt}
\tablehead{
\colhead{Time$^{\mbox{\scriptsize a}}$}  &  \colhead{(UTC)}   &   \colhead{Aug. 04 15:00--16:37} &   \colhead{Aug. 04 17:15--18:52} &   \colhead{Aug. 06 16:22--18:28} &   \colhead{Aug. 18 15:15--16:52} 
}
\startdata
Observation &   & Ku band epoch 1    &   Ku band epoch 2  & X band &   Ku band epoch 3 \\
Frequency   &   (GHz)   &   12--18   &   12--18   &   8--12    &   12--18   \\
{\it uv}-distance   &   (meters)    & 730--36550 &   $\cdots$    &   600--30840   &   525--28340   \\
Synthesized beam$^{\mbox{\scriptsize b}}$    &  (\beam; BPA) &   0\farcs18$\times$0\farcs13; 33$^{\circ}$    & $\cdots$  &   0\farcs40$\times$0\farcs26; 65$^{\circ}$    &    0\farcs16$\times$0\farcs14; 25$^{\circ}$   \\
RMS noise$^{\mbox{\scriptsize b}}$   &   (\ujyperbeam)   &   5.8 &   $\cdots$    &   5.3 &   4.7 \\
Peak intensity$^{\mbox{\scriptsize b,c}}$  &   (\ujyperbeam)   &   26  &   $\cdots$    &   21  &   27  \\
Flux density$^{\mbox{\scriptsize b,c}}$    &   ($\mu$Jy)   &   120 &   $\cdots$    &   48  &   90  \\
Gain calibrator flux$^{\mbox{\scriptsize d,f}}$    &   (Jy)    &   0.63    &   $\cdots$    &   0.62    &   0.64    \\
BP calibrator   flux$^{\mbox{\scriptsize e,f}}$    &   (Jy)    &   35  &   $\cdots$    &   38  &   35  \\
\enddata
\tablenotetext{a}{Time range for the target source loop. All observations were taken in 2019.}\vspace{-0.2cm}
\tablenotetext{b}{Measured from naturally weighed images.}\vspace{-0.2cm}
\tablenotetext{c}{Measurements from the target source DM~Tau.}\vspace{-0.2cm}
\tablenotetext{d}{J0449+1121.} \vspace{-0.2cm}
\tablenotetext{e}{J0319+4130.}\vspace{-0.2cm}
\tablenotetext{f}{A $\sim$5\% nominal absolute flux calibration accuracy (which is higher than thermal noise) has to be assumed according to the official documentation (\href{https://science.nrao.edu/facilities/vla/docs/manuals/oss/performance/fdscale}{https://science.nrao.edu/facilities/vla/docs/manuals/oss/performance/fdscale}).}
\end{deluxetable*}

Pre-main sequence (PMS) stars younger than a few Myr still retain a significant fraction of angular momentum at birth and thus typically have short rotation periods of a few days \citep{Edwards1993}.
These young PMS stars are prone to generate magnetic structures via the geodynamo effects.
As a consequence, fast rotating PMS stars are known to be associated with powerful magnetic activities that are orders of magnitude more intense than in the main sequence stars. 
The dynamical evolution of the magnetic structures (e.g., \citealt{Favata2005}) and the interplay between the magnetic structures, the PMS stars, and the natal disks can be manifested by time variations in the photometric magnitudes over wide wavelength ranges.
For instance, the surfaces of the young PMS stars are often populated with giant cold spots which are the bases of magnetic loops.
The presence of cold spots leads to periodic variations of photometric magnitudes (e.g., \citealt{Kospal2018}).
In addition, the magnetic loops may inflate stochastically, leading to shocks at 10$^{6}$--10$^{7}$ K.
The high-temperature shocks make PMS stars ubiquitous thermal X-ray emitters, which typically vary from hours to days or longer.
Moreover, when the magnetic loops are coupled with the inner protoplanetary disks, it can also lead to episodic accretion shocks and may excite disk warps that cast time-variable extinction
(see \citealt{Feigelson1999,Armitage2016ApJ...833L..15A} and references therein).
These phenomena now can be investigated in unprecedented detail thanks to the short-cadence and precise photometric monitoring observations from CoRoT \citep{Baglin2006}, Kepler/K2 \citep{Howell2014}, and TESS \citep{Ricker2015} space missions (e.g., \citealt{Cody2022}).
The outstanding questions to be addressed with these studies include how the high-energy magnetic flares of the young PMS stars affect the (bio)chemistry in the natal protoplanetary disks (e.g., \citealt{Henning2010,Semenov2011,Teague2015,Cleeves2017,Espaillat2022ApJ...924..104E}), the formation of chondrite and planetesimals (e.g., \citealt{Feigelson1999}), the disk mass dispersal (e.g., \citealt{Owen2010,Pascucci2014ApJ...795....1P}), the migration history of the (proto)planetary systems (e.g., \citealt{Monsch2019}), and the habitability of the newly formed planets. 

In addition, radio observations may probe the properties of the ionized photoevaporation wind which have been less explored.
\citet{Pascucci2012} suggested that the observations of free-free emission at centimeter bands can test the disk photoevaporation models, which is linked to disk mass dispersal. 
In a case study, it has been reported that the X-ray and EUV-driven photoevaporation is consistent with the results of radio observations \citep{Owen2013} although it may not be general (c.f., \citealt{Galvan2014A&A...570L...9G,Coutens2019A&A...631A..58C}).

% \citet{Guarcello2019} recently reported that anti-correlated and positively correlated X-ray and optical light curves, which have been found in emissions from sunspots and faculae regions of the Sun were also reported to have been detected in several PMS stars. 
% In the meantime, PMS stars have been observed frequently as sites where star formation and planet formation may be ongoing.
% ALMA long baseline observations have ubiquitously resolved dusty disk structures such as gaps and asymmetric structures \citep{Andrews2018}. 

To improve the understanding of the interplay between the high-energy emission of the PMS star and the protoplanetary disk, we have carried out the focused observational case study towards the M1 type PMS star, DM~Tau ($M_{*}\sim$0.5 $M_{\odot}$, $d\sim$144 pc; \citealt{gaia2020}).
The M-type PMS stars tend to be convective and may form large magnetic dipoles, making the problem simpler and more comprehensive than the higher mass stars that may be associated with rather complicated magnetic structures.
The previous Submillimeter Array (SMA) 0.88 mm continuum image towards the DM~Tau disk marginally resolved a dust ring at $\sim$19 au radii \citep{Andrews2011} that is $\sim$30$^{\circ}$--35$^{\circ}$ inclined (\citealt{Kudo2018,Hashimoto2021}).
The disk of DM~Tau provides an opportunity to trace the ionized gas structures in the disk using spatially resolved radio observations.
The follow-up Atacama Large Millimeter/submillimeter Array (ALMA) long-baseline observations further spatially resolved an  $\sim$3 AU scales inner disk that is encircled by the dust ring \citep{Kudo2018,Hashimoto2021}.
%\textbf{The inner disk is located around $\sim 3$ au and the outer disk around $\sim 20$ au.}
In the ALMA image, the inner disk presents an inner cavity, which is consistent with what was indicated by the analysis of the optical and infrared spectral energy distribution (SED; \citealt{Calvet2005}).
The rich (sub-)structures in the DM~Tau disk could be readily explained by planet-disk interactions \citep{Kanagawa2015,Dong2017ApJ...835..146D}.
The studies on this system will potentially provide us a view of how the ionizing photons emanating from the PMS star impact the planet-forming disks.

% With the aim of investigating the relationship between cool spots of DM~Tau and its transitional disk, we analyzed the high cadence optical photometric monitoring data observed by K2 and ground-based optical photometric monitoring data to characterize cold spots on DM~Tau.
% We also investigate the relationship between the variability seen in optical and the spatially resolved extended radio emission observed by JVLA. 
% Also, recent ALMA observations indicate that the outer disk has two asymmetries and that the emissions inside the inner disk can be caused by small icy planets \citep{Hashimoto2021}.

We have carried out optical and near-infrared monitoring photometric observations towards DM~Tau in early and late 2019.
We jointly analyzed these observations with the K2 \citep{Howell2014} light curves. 
We identified periodic variations from the K2 optical light curves, and then based on the multi-color photometry to deduce the origin of optical/infrared variabilities. 

In addition, we performed three epochs of radio observations using the NRAO\footnote{The National Radio Astronomy Observatory is a facility of the National Science Foundation operated under cooperative agreement by Associated
Universities, Inc} Karl G. Jansky Very Large Array (JVLA) to spatially resolve the distribution of ionized gas in the DM~Tau disk.
The details of our observations and data reductions are introduced in Sections \ref{sec:obs}.
Section \ref{sec:results} presents an examination of the periodicity in the optical and near-infrared light curves, which is compared with the observed variability at radio wavelength.
We attribute the optical/infrared variability to the presence of cold spot(s).
Our simple model of stellar spot to compare with the observations is outline in Appendix \ref{appendix:spotmodel}.
The joint interpretation of the present and previously published optical, radio, and (sub)millimter data is discussed in Section \ref{sec:discussion}.
Our conclusion is given in Section \ref{sec:conclusion}.

\begin{figure*}
    \hspace{-1.2cm}
    \begin{tabular}{cc}%{ p{8.8cm} p{8.8cm} }
        \includegraphics[width=9cm]{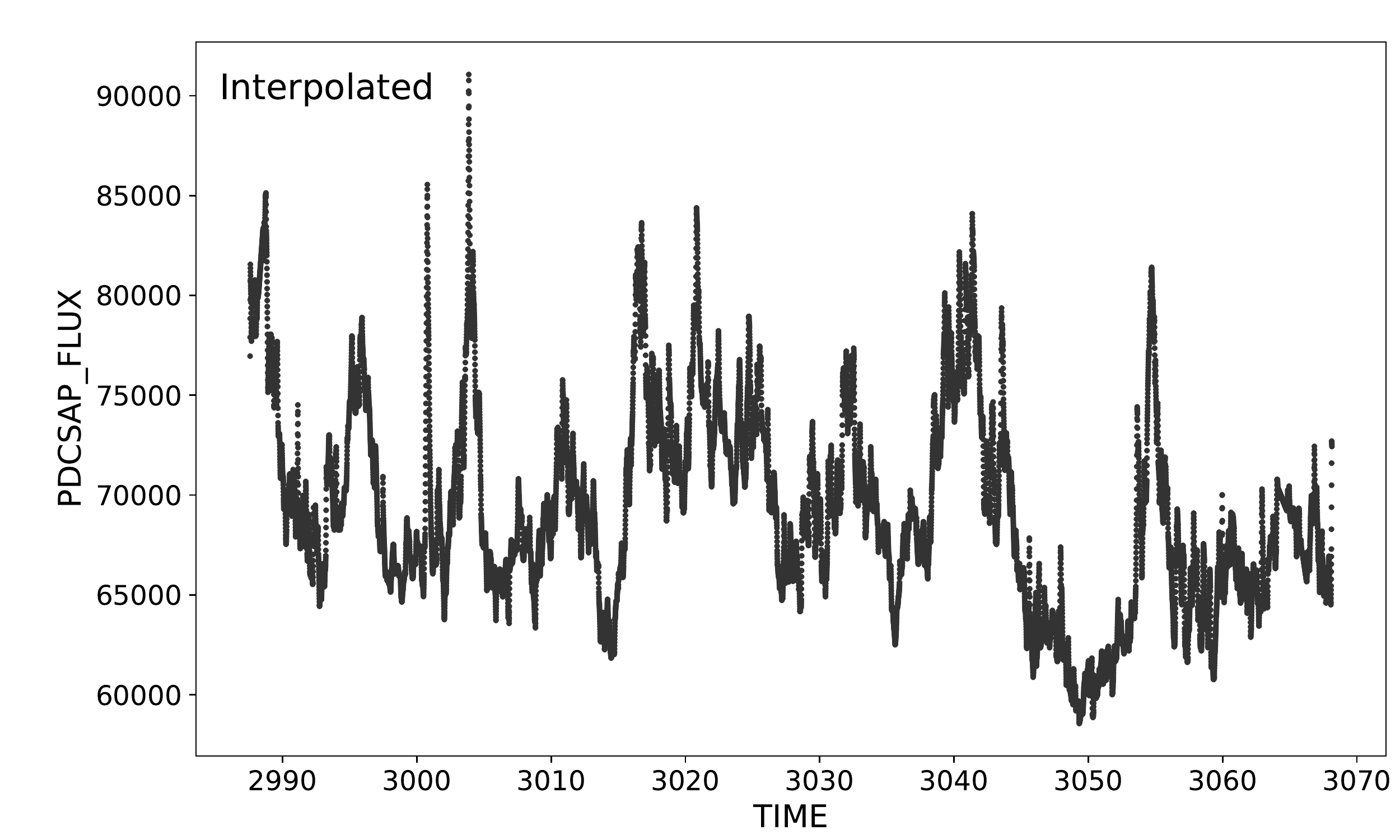}  &
         \includegraphics[width=9cm]{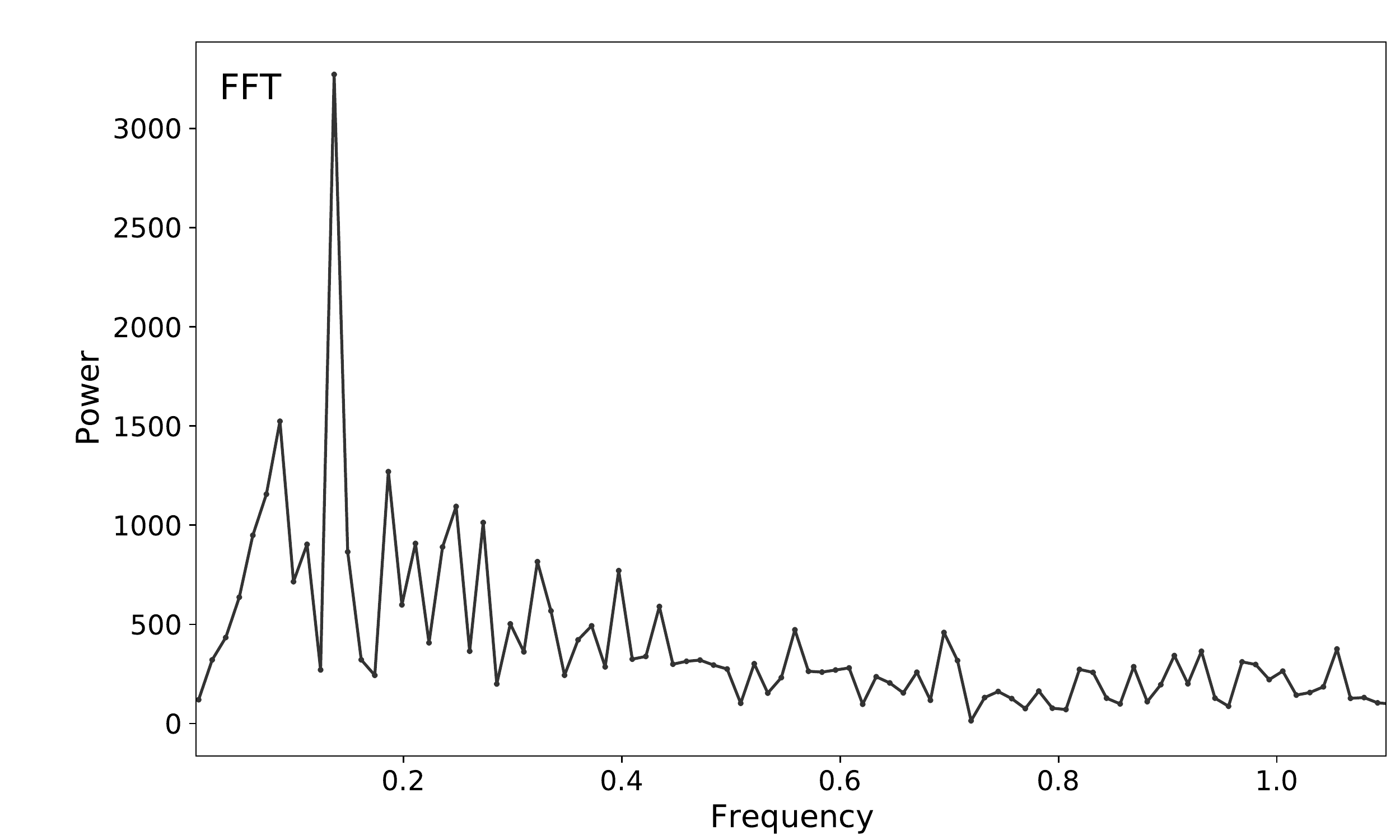} \\
         \includegraphics[width=9cm]{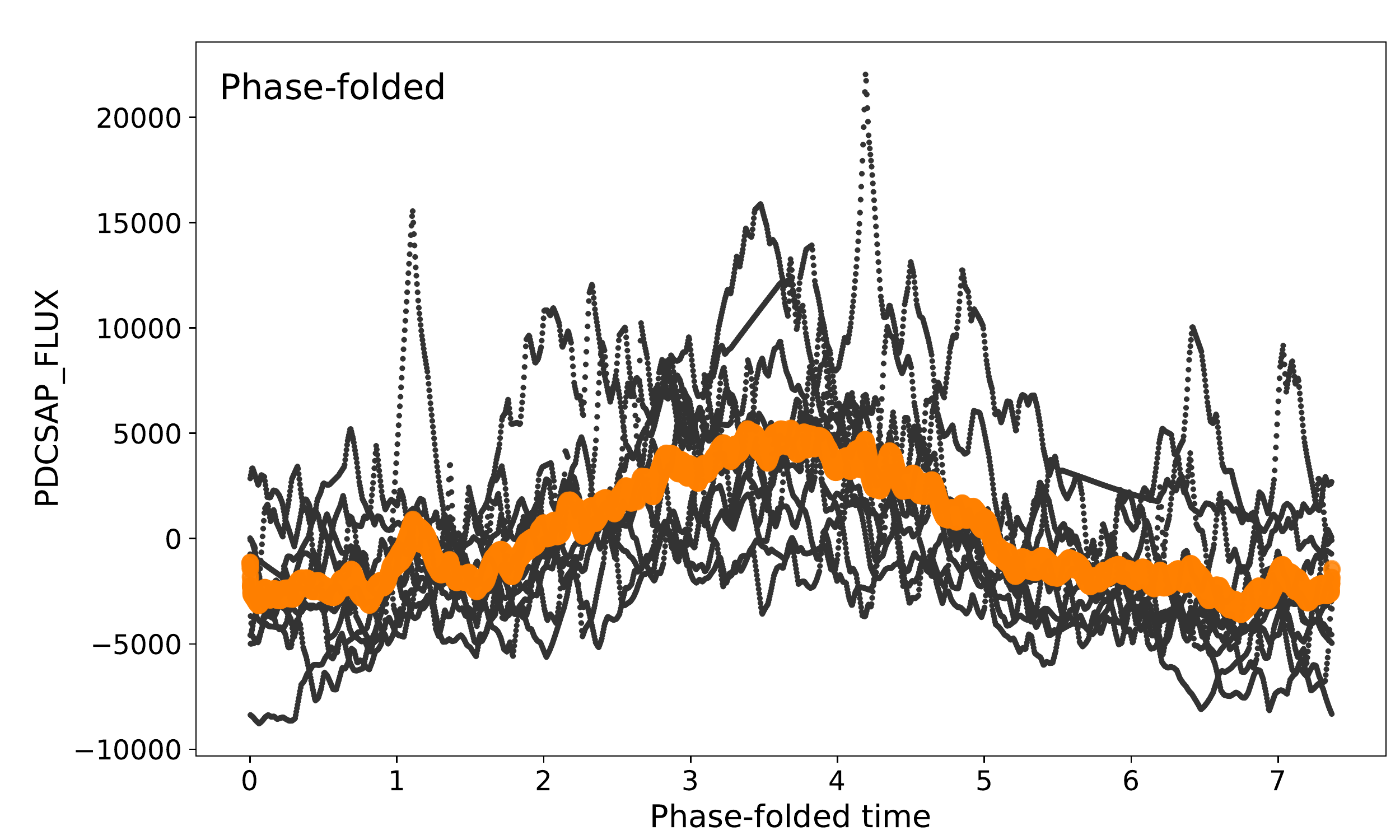} &
          \includegraphics[width=9cm]{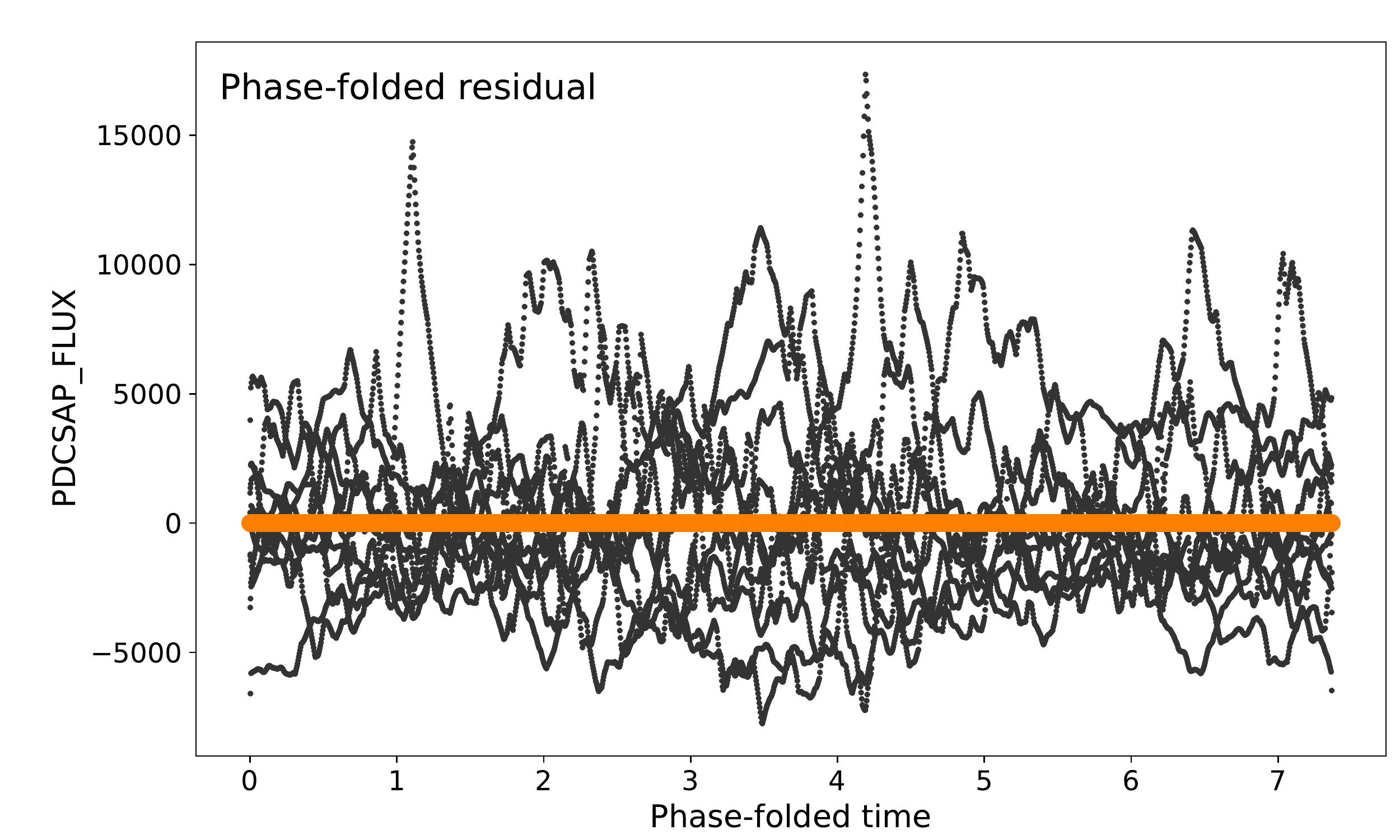} \\
    \end{tabular}
    \caption{The K2 data on DM~Tau. {\it Top left :--} K2 light curve which was interpolated onto the time grids with 6 minutes interval. {\it Top right :--} Power spectrum, where the vertical axis is presented in arbitrary units.
    %Inset shows a close-up power spectrum in frequency below 0.5.
    {\it Bottom left :--} The light curve phase-folded with a period of 7.364 days (black), over-plotted with the boxcar-smoothed curve by a 0.1-days windows (orange). {\it Bottom right :--} Residual evaluated by subtracting the orange curve from the black curve in the bottom left panel. For the detailed data manipulation see Section \ref{sub:period}.}
    \vspace{0.5cm}
    \label{fig:K2data}
\end{figure*}

\section{Observations} \label{sec:obs}

\subsection{JVLA observations} \label{sub:obsJVLA}
We have performed the JVLA observations towards DM~Tau at X (8---12 GHz) and Ku (12--18 GHz) bands in the A array configuration in 2019 August.
The observing dates, frequency ranges, and {\it uv}-distance ranges (after data flagging) are summarized in Table \ref{tab:JVLAobs}.
All epochs of observations used J0521+1638 (3C138), J0319+4130 (3C84), and J0449+1121 for absolute flux, passband, and complex grain calibrations, respectively.

The second epochs of observations on Aug.~04 (UTC 17:15--18:52) cannot be calibrated due to that the weather condition degraded with time significantly.
For the rest of the observations, we manually calibrated the JVLA data following the standard strategy using the Common Astronomy Software Applications \citep[CASA;][]{McMullin2007} package release 5.6.0.
After implementing the antenna position corrections, weather information, gain-elevation curve, and opacity model, we bootstrapped delay fitting and passband calibrations, and then
performed complex gain calibration. 
We applied the absolute flux reference to our complex gain solutions, and then applied all derived solution tables to the target source.
The absolute fluxes of the flux calibrator 3C138 were referenced from the  Perley-Butler 2017 standards \citep{Perley2017}.

We performed the first order (i.e., nterm$=$2) multi-frequency synthesis imaging \citep{Cornwell2008MultiscaleClean,Rau2011MultiscaleClean} using the CASA task {\tt tclean}.
When assessing the data quality, we discovered that the morphology of the radio intensity distribution varies from epoch to epoch.
Therefore, instead of jointly imaging all epochs of observational data, we imaged individual of them separately, using the Briggs Robust$=$2 weighting to maximize the signal to noise ratios.
The achieved synthesized beams and root-mean-square (rms) noise levels are summarized in Table \ref{tab:JVLAobs}.
We additionally produced the Robust$=$0 weighted X band image (\beam=0\farcs30$\times$0\farcs19; P.A.$=$69$^{\circ}$; rms$=$7 $\mu$Jy\,beam$^{-1}$) of which the synthesized beam size is closer to those of the Robust$=$2 weighted Ku band images.

% \todo{It looks like the Robust=2 weighed X band image have too poor angular resolution. We should use the Robust=0 weighted one. I just shared the FITS images to you (also the ALMA FITS images). I have implemented the proper motion correction to the ALMA image such that the disk appears at where is should be during the time epoch of JVLA observations. We should derive the central position from this ALMA image instead of quoting the center from Kudo san's paper.}

\begin{figure*}
    \hspace{-1cm}
    \begin{tabular}{cc}
         \includegraphics[width=9cm]{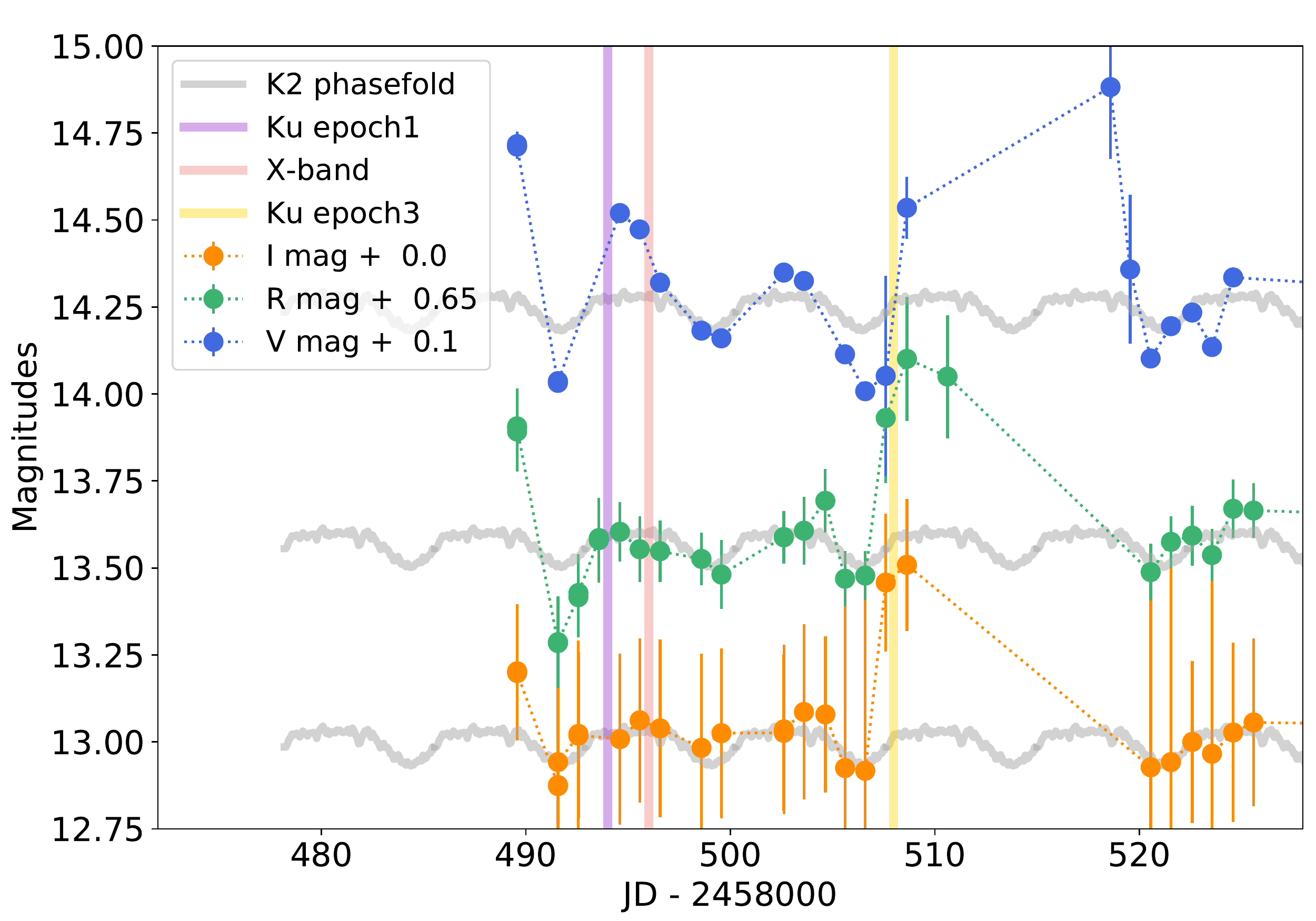} & 
         \includegraphics[width=9cm]{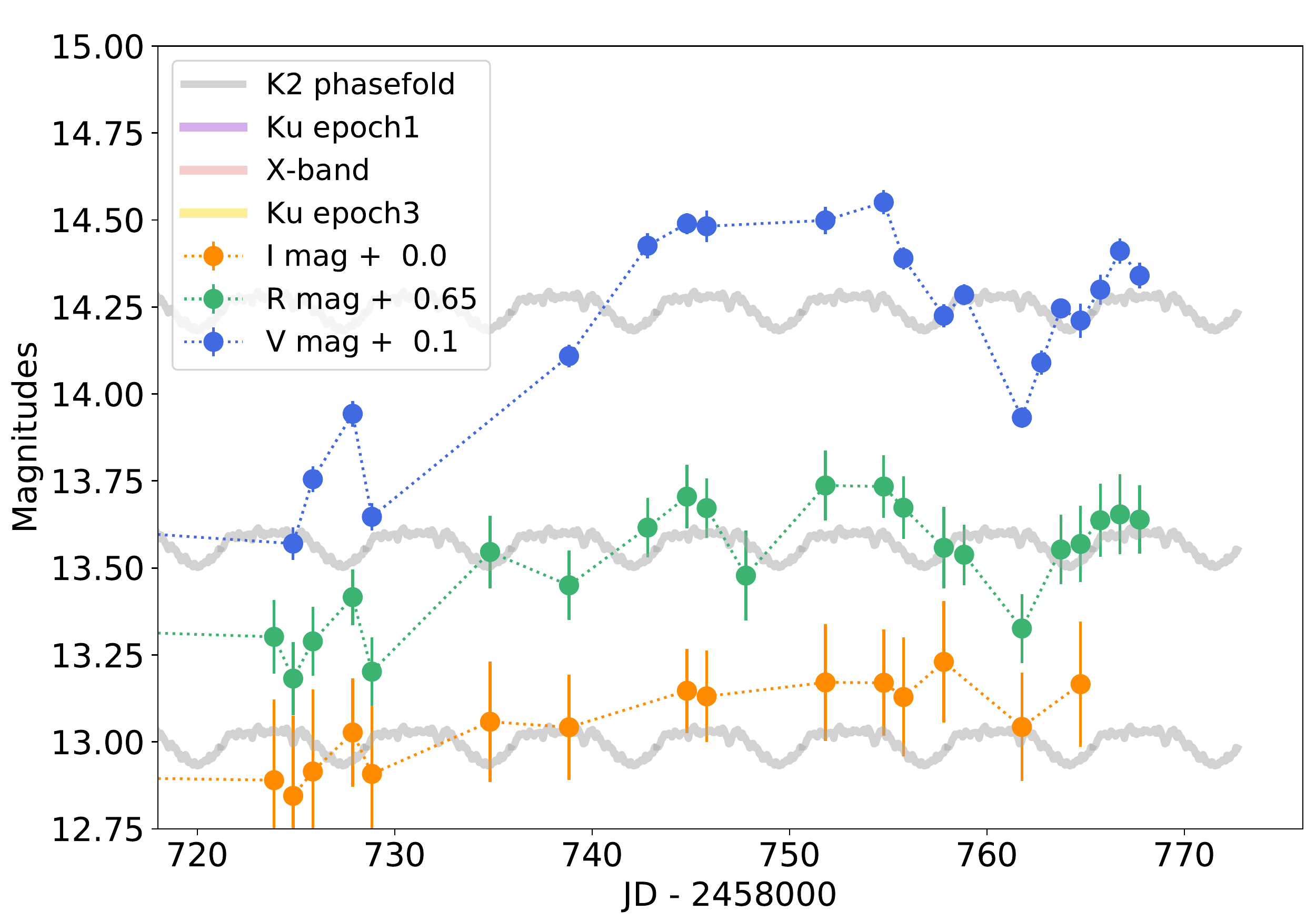} \\ 
    \end{tabular}
    \caption{Optical and near-infrared monitoring observations towards DM~Tau. Green, orange, and cyan symbols present our V, R, and I band monitoring observations that were taken in 2019 (Section \ref{sub:obsBVI}). Left and right panels show the observations were taken in early and late 2019, respectively, which were separated by a time period when DM~Tau is not visible from our observatory at night time.
    Gray curves show replications of the phase-folded K2 and 0.1-day boxcar-smoothed light curve (Section \ref{sub:period}), which are vertically offset to permit easier comparison with the V, R, and I bands data. The horizontal axis shows the Julian day offsets from reference 2458000.
    Three vertical lines in the left panel indicate the epochs of the JVLA observations which have been shifted by -28 periods for easier comparison with the optical monitoring.}
    
    \vspace{0.5cm}
    \label{fig:monitoring}
\end{figure*}

\subsection{VRI bands monitoring in 2019} \label{sub:obsBVI}

The multi-color optical monitoring of DM\,Tau was carried out with the 0.6-m robotic PROMPT-8 telescope of Thai National Astronomical Institute at Sierra-Tololo Inter American Observatory (CTIO), Chile ($70.805^\circ$ W, $30.168^\circ$ S). 
The observations were collected during 48 nights and span one year between January 12, 2019, to December 22, 2019, with a 2048$\times$2048 pixels size  CCD camera with a scale of 0.624 arcsec/pixel. The  V, R, and I filters were used and the typical exposures were 100, 80, and 80 seconds respectively.

After performing dark-image subtraction and flat-fielding correction, we produced preview images for visually picking out the epochs with poor focuses, poor source tracking, very bright sky, the noisy observations where DM~Tau cannot be significantly detected, or those which were influenced by other artifacts.
%\todo{There were 12 epochs of observations which were not analyzed, due to missing coordinate headers. Needs Yuka to work on re-constructing the coordinate headers.}
From the remaining images, we identified comparison stars that are within $\pm$1 magnitude from DM~Tau (i.e., counts are lower than 2.5 times that of DM~Tau and higher than $1/2.5$ times that of DM~Tau).
We then performed Gaussian fittings to the comparison stars to derive the full width at half maximum (FWHM) of the point spread functions (PSFs) for individual epochs of the observations and then performed aperture photometric measurements on the comparison stars.
We made a query to the Vizier database to obtain (if available) the apparent magnitudes of the comparison stars; those which do not have record in Vizier were discarded.
Due to the artifacts in the images, and due to that the field-of-views were not exactly the same from epoch to epoch in our robotic monitoring, the numbers of usable comparison stars are not a fixed number.
Normally, individual epochs of observations covered $\sim$20--30 usable comparing stars.

Finally, for individual epochs of observations, we adopted the median PSF-FWHM derived from the comparison stars to perform aperture photometric measurements on DM~Tau, and then calibrate the CCD counts of DM~Tau to magnitudes by referencing the $\sim$20--30 comparison stars.
We adopted the biweighted mean and biweighted standard deviation of the $\sim$20--30 referenced magnitudes as the magnitude and magnitude error (1-$\sigma$) of DM~Tau at that epoch.
Our magnitude errors therefore not only reflect the thermal noise, but also the systematic effects including the time-varying artifacts in the images (e.g., imperfect flat fielding) and also that some comparison stars might be variables.

\begin{figure*}
    \hspace{0cm}
    \begin{tabular}{cc}%{ p{6cm} p{6cm} }
         \includegraphics[height=16cm]{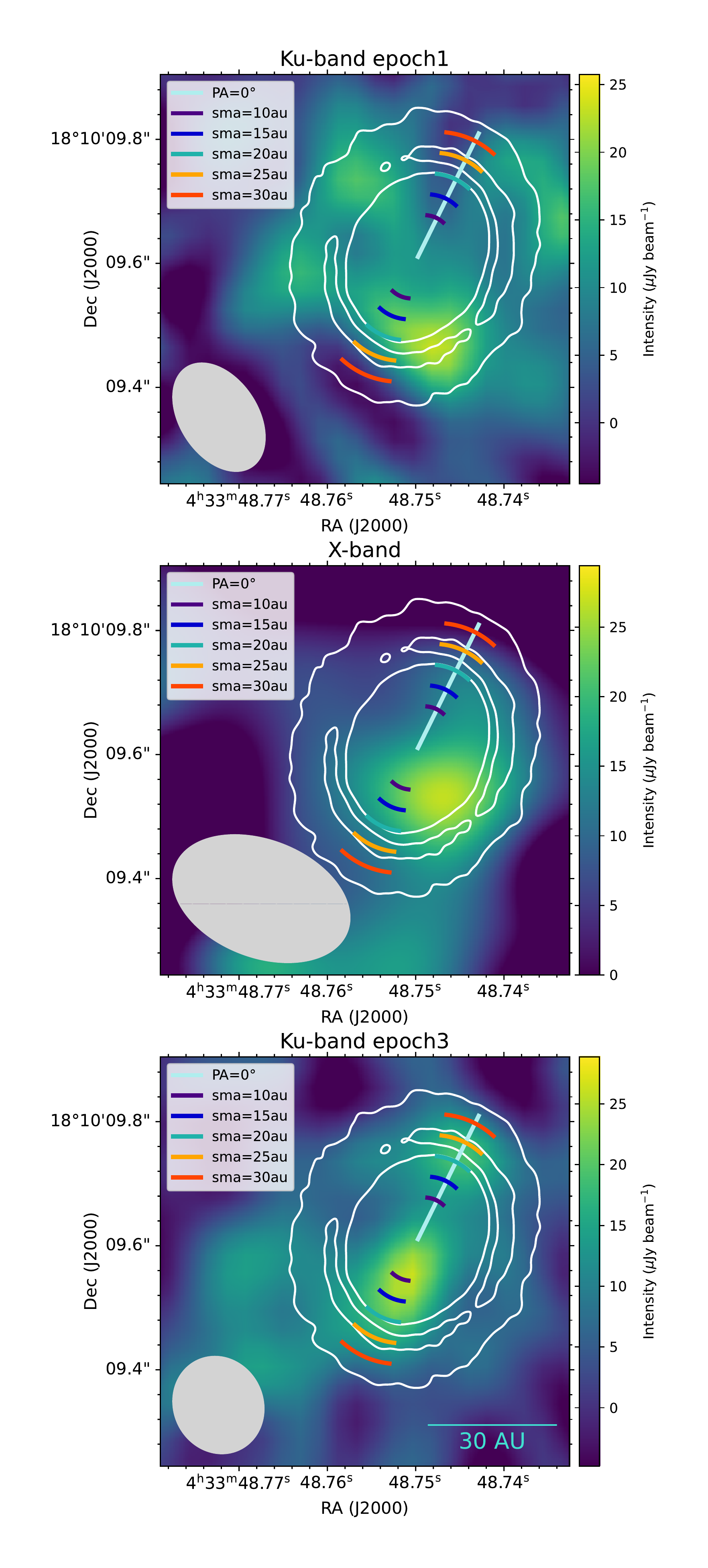} & 
         \includegraphics[height=16cm]{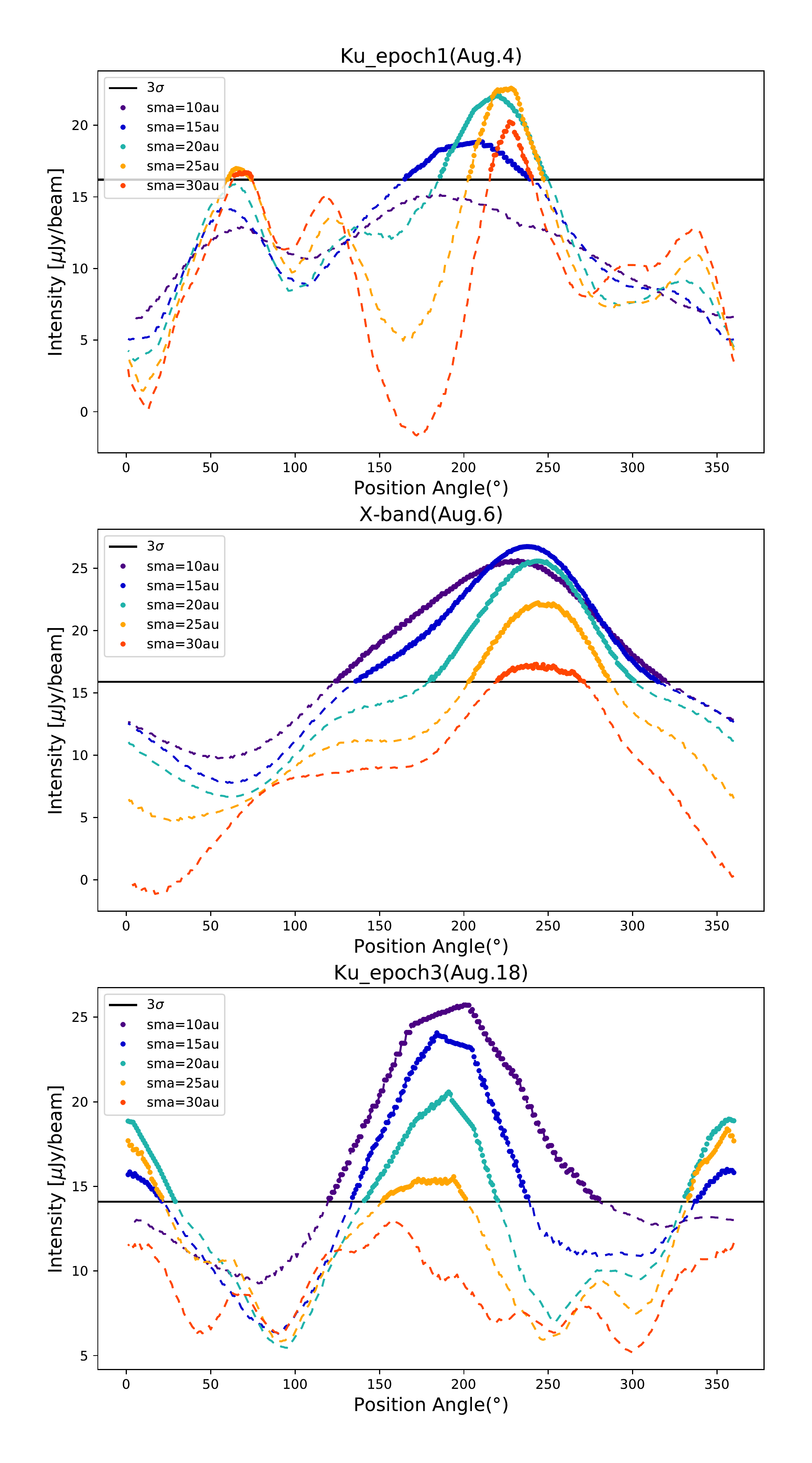}
    \end{tabular}
  \caption{
  {\it Left :-} 
  The JVLA images (color; Table \ref{tab:JVLAobs}), are overplotted with the ALMA 1.3 mm continuum image (\citealt{Hashimoto2021}). Contours are [200, 400] $\mu$Jy\,beam$^{-1}$ (\beam=0\farcs038$\times$0\farcs026; P.A.=33$^{\circ}$).
  From top to bottom rows show the Robust$=$2 weighted Ku band (12--18 GHz) image taken on August 4, the Robust$=$0 weighted X band (8--12 GHz) image taken on August 6, and the Robust$=$2 weighted Ku band image taken on August 18.
  Color-coded arcs indicate the [10, 15, 20, 25, 30] au radii.
  The lightblue line indicates the major axis of the DM~Tau disk. The filled ellipse denotes the JVLA synthesized beam.
  {\it Right :-} 
  Azimuthal profiles of the radio intensity derived from the images presented in the left panel. In these plots, the zero position angle is defined along the major axis of the disk. Black lines indicate 3-$\sigma$ detection limits.  }
  \vspace{0.2cm}
  \label{fig:jvla_intensity}
\end{figure*}

\subsection{K2 light curve} \label{sub:K2}

The K2 long-cadence time series observations on DM~Tau were acquired in 2017 during Campaign 13 (March 8--May 27). 
We retrieved the lightcurve from the Milkulski Archive for Space Telescopes (MAST\footnote{\url{http://archive.stsci.edu/k2/data_search/search.php}}) and removed the data points with bad quality flags.
%The K2 data provides the best signal-to-noise ratio and the smallest time cadence for examining the periodicity of the time variation.
%We downloaded the observations on DM~Tau from the K2 data archive (\todo{Check which is the proper reference.}).
%\todo{describe more about the specific K2 campaign which covered our targetsource .}
%There are several techniques have been developed to correct systematic errors so we used the data values for EVEREST(\citet{Luger2016},\citet{Luger2018}). 
Afterward, we removed outliers that are offset from the median light curve by more than 5-$\sigma$.
% \question{Do you mean: We tried producing the median light curves by median smoothing with a 1-day and 0.5-day smoothing. How did you produce the median light curve with no smoothing?}
% \textcolor{cyan}{
To avoid excluding the signals from the target, we have tried setting the light curves by median smoothing on ~1-day, 0.5-day, and no smoothing.
% }
We converted the PDCSAP flux values ($F_{\mbox{\tiny K2}}$) of the K2 data to magnitudes ($m_{\mbox{\tiny K2}}$) by
\begin{equation}
    m_{\mbox{\tiny K2}} = m_{\mbox{\tiny K2}}^{0} - 2.5 \log_{10}  F_{\mbox{\tiny K2}},
\end{equation}
where $m_{\mbox{\tiny K2}}^{0}$ can be an arbitrary magnitude offset.

% \textcolor{cyan}{Multi-band photometry showed the color changes related to the flux variations.} % Baobab: this should be in the result section instead of observation section

% \todo{Double check whether there are variables in our comparison star catalogues.}

%\todo{How to choose comparison stars?}

%\begin{figure*}
%    \hspace{-1cm}
%    \begin{tabular}{cc}
%         \includegraphics[width=9cm]{CMD_vr.pdf} & 
%         \includegraphics[width=9cm]{CMD_vi.pdf} \\ 
%    \end{tabular}
%    \caption{V vs.(V-R) (left panel) and V vs.(V-I) (right panel) color-magnitude diagrams of the system to determine interstellar extinction. The black line in each panel is a linear regression fit to the color slope. The color code reflects the Julian Date of observations.
%    }
%    \label{fig:CMD}
%\end{figure*}

\section{Results} \label{sec:results}

\subsection{Optical and infrared photometric monitoring} \label{sub:period}

Following the approach of \citet{Cody2014}, we first interpolated the K2 data onto regularly spaced time grids with 6 minutes intervals (Figure \ref{fig:K2data}, top left).
We then boxcar smoothed the interpolated light curve using a 10-days window, and then subtracted the boxcar-smoothed light curve from the interpolated one to suppress the variability on $\gtrsim$15--20 days timescales.
Afterward, we performed Fast Fourier Transform (FFT) to derive the power spectrum of the light curve and then made an initial identification of the characteristic variational period ($\tilde{P}$) based on the peak of the power spectrum (Figure \ref{fig:K2data}, top right).
We phase-folded the light curve according to $\tilde{P}$, and then followed the convention of \citet{Cody2018} to evaluate the periodicity metric Q which is defined as
\begin{equation}
    \frac{r_{\mbox{\tiny resid}}^2 } { r_{\mbox{\tiny raw}}^2},
\end{equation}
where $r_{\mbox{\scriptsize raw}}$ is the rms of the light curve, and $r_{\mbox{\scriptsize resid}}$ is the rms evaluated after subtracting a boxcar-smoothed light curve from the phase-folded one (Figure \ref{fig:K2data}, bottom left).
In our case, the thermal noise of the light curve ($\sigma$ in \citealt{Cody2018}) is negligibly small, and therefore the error of Q is dominated by the uncertainty of $\tilde{P}$.
We varied the value of $\tilde{P}$ from the aforementioned initial identification to minimize Q.
We found a minimum of Q$=$0.63 when using a period of 7.364 days, which we regard as the best-fit variational period of DM~Tau.
Our derived period is consistent with what was reported in \citet{Rebull2020}.
We visually inspected the phase-folded light curves and the residuals after subtracting the boxcar-smoothed light curves (Figure \ref{fig:K2data}, bottom right) to ensure the correct convergence of the period and Q.
We note that in addition to the $7.364 \pm 0.042$ days periodic variations, the K2 light curve of DM~Tau also occasionally presents flares of higher fluxes and shorter duration.

%%We also followed \citet{Cody2014} to derive the asymmetric metric
%%\begin{equation}
%%    M = (<d_{10\%}>-d_{\mbox{\scriptsize med}})/ \sigma_d,
%%\end{equation}
%%where $\sigma_d$ is the rms of the light curve \todo{check what does this mean in Cody+18}, $d_{\mbox{\scriptsize med}}$ is the median of the light curve after subtracting the light curve which was boxcar smoothed with a 2-hours window, $<d_{10\%}>$ is mean of all data at the top and bottom decile of the same light curve after subtraction.
%%We found M$\sim$0.12. \todo{benchmark M from some examples in Cody+18.}
The derived Q value indicates that DM~Tau may be regarded as a quasi-periodic variable at optical wavelengths according to the classification of \citet{Cody2018}.
We note that although \citet{Cody2022} reported the quasi-periodicity of the DM~Tau light curve, they did not formally classify DM~Tau as a quasi-periodic variable due to that it presented rich optical bursts. Instead, they classified DM~Tau as a burster. 
% \textcolor{cyan}{
%Also, \citet{Cody2022} reported that DM Tau is evaluated as a quasi-periodic in terms of its period and Q, but they do not consider DM Tau quasi-periodic because of the light curve is dominated by bursting behavior.  
%}

Figure \ref{fig:monitoring} shows the V, R, and I bands monitoring results (Section \ref{sub:obsBVI}), over-plotted with the replications of the phase-folded K2 light curve.
The V, R, and I magnitudes are clearly variable.
%In general, \textcolor{cyan}{the temperature difference between the cold spots and non-spotted photosphere is a few hundred Kelvin.
%Due to the feature, the cold spots variations are wavelength dependent, with larger variations appearing in the shorter wavelength V-band.}
%the magnitude variations at V band are higher than those in R and I bands.
During the Julian days (JD) 2458490--2458508, the V band monitoring clearly resolved the periodic variation of which the time and periods are consistent with those of the K2 light curve. At the same time, the R band shows a similar variation to the V band, although the error bar is larger.
In addition, DM~Tau occasionally presents a variation of $\sim\pm$0.5 magnitudes on timescales shorter and longer than 7.364 days.
For example, a long-term variation of $\gtrsim$20 days timescale can be seen from the right panel of Figure \ref{fig:monitoring}.
The observed OIR light curves can show offsets from a perfect periodic light curve due to stochastic flares (e.g., \citealt{Chang2018ApJ...867...78C,Hawley2014ApJ...797..121H}) or the switching of accretion column (e.g., \citealt{Cemeljic2020MNRAS.491.1057C}).

% \textbf{The lightcurve variability of PMS stars may be mainly due to accretion hotspots, cool spots, and extinction; DM Tau also shows complex variability in the K2 lightcurve, suggesting a mixture of multiple variations. 
% Among these variations, cool spots show periodicity, and based on \citep{Grankin2007} results, we consider the periodic variations obtained to be cool spot origin.}

If the 7.364 days periodic variation of the DM~Tau is stationary, then our JVLA X band observations were taken close to the minimum of this periodic optical modulation; the first epoch of Ku band observations were taken slightly closer to the maximum while the second Ku band observations were taken further closer to the maximum.
%The JVLA images are introduced in the following Section \ref{sub:JVLAimage}.

\begin{deluxetable*}{ l c c c  }
\tablecaption{Input parameters and results of model fitting \label{tab:input_params}}
\tabletypesize{\scriptsize}
\tablecolumns{5}10.24022
%\tablenum{1}
%\tablewidth{10pt}
\tablehead{
\colhead{  Free parameter} &&&\\
}
\startdata
Parameter  &  Fitting parameter range            &  fitting result  &  Comments \\\hline
   $R_{\star}$[$R_\odot$]                 & $> 0 $         & $1.986^{+0.054}_{-0.014}$       & stellar radius    \\
   $R_{\mathrm{spot}}$[$\mbox{R}_{\star}$] &  0 to 1       &  $0.999989^{+0.000010}_{-0.000065}$ &  spot size  \\
   $T_\mathrm{hot}$[K]                    & 3500 to 4500  & $3513.99^{+4.42}_{-13.98}$      & non-spotted temperature on the photoshpere \\
   $T_\mathrm{spot}$[K]              & 2500 to 3500  & $2505.78^{+3.95}_{-5.77}$       & spot temperature \\
   \textit{long}[$^\circ$]                           & -180 to 180   &  $10.24^{+0.66}_{-1.25}$        & spot longitude \\
   \textit{lat}[$^\circ$]                            & -90 to 90     &  $84.29^{+1.45}_{-0.63}$        & spot latitude    \\ \hline \hline
Fixed parameter &&&\\
    & & & \\ \hline
   Parameter                          &  Fiducial value  &  &  Comments\\\hline
    $\mbox{P}_{\mathrm{rot}}$[day]    &  $7.367^{\textit{a}}$  &  &  rotation period \\
    d [pc]                            & $144.048^{\textit{b}}$ &  &  distance \\
    $\log g$                          & $3.9849^{\textit{c}}$  &  &  surface density \\ 
    {[Fe/H]}                      & $0.14^{\textit{d}}$    &  &  metallicity \\
    $A_{\mbox{\scriptsize v}}$    &  $1.5^{\textit{e}}$    &  &   interstellar extinction \\ 
    $v \mathrm{sin} i$ [km/s]     &  $10^{\textit{f}}$  &  & projected rotational velocity \\
   $N^{\mbox{\scriptsize grid}}$  & 300              &  &  $N^{\mbox{\scriptsize grid}}\times N^{\mbox{\scriptsize grid}}$ resolution elements\\
\enddata
  \tablenotetext{a}{K2 analysis}\vspace{-0.25cm}
  \tablenotetext{b}{\citealt{gaia2020}}\vspace{-0.25cm} 
  \tablenotetext{c}{\citealt{Woitke2019}}\vspace{-0.25cm}
  \tablenotetext{d}{\citealt{Lastennet1999}}\vspace{-0.25cm}
  \tablenotetext{e}{\citealt{Grankin2007}}\vspace{-0.25cm}
  \tablenotetext{f}{\citet{Hartmann1989}}
\end{deluxetable*}

\subsection{Centimeter band images}\label{sub:JVLAimage}
The radio flux densities we detected (Table \ref{tab:JVLAobs}) are comparable with those of the previously detected Class II young stellar objects (YSOs) although a large number of Class II YSOs remain undetected at radio bands (c.f., \citealt{Liu2014ApJ...780..155L,Galvan2014A&A...570L...9G,Dzib2015ApJ...801...91D,Coutens2019A&A...631A..58C}).
Figure \ref{fig:jvla_intensity} shows the JVLA Robust$=$0 weighted 8--12 GHz (X band) and Robust$=$2 weighted 12--18 GHz (Ku band) images. 
Each observation was performed on a different day (Table \ref{tab:JVLAobs}).
The spatially resolved extended radio emission does not coincide with the 1.3 mm emission and the location of the host PMS star (Figure \ref{fig:jvla_intensity}) and therefore can only be interpreted by free-free emission from ionized gas. 
%\todo{A journal paper cannot publish an image that is a reproduction of what was published before. Remove Figure 3 and overplot the contours in the left panel of Figure 4. Maybe you can change the colorful ellipses to short arcs (e.g., $\pm$15$^{\circ}$ offset from the cyan bars) to avoid cluttering presentation.}
The X band intensity distribution can be described by a compact bright emission knot in the southwest and a diffuse ionized crescent that occupies the southwest side of the millimeter cavity.
The first epoch of Ku band image significantly detected a knot in the south. 
The third epoch of Ku band image detected a bright knot in the southeast; there might be a fainter knot in the northwest.

It appears that the position angles of the radio intensity peaks vary from epoch to epoch. 
The spatial variation of the centimeter band emission can be more clearly seen in the azimuthal intensity profiles of the JVLA images (Figure \ref{fig:jvla_intensity}).
The right panel of Figure \ref{fig:jvla_intensity} shows the azimuthal radio intensity profiles as functions of radii.
We defined the centroid position angles of the radio emission knots based on performing Gaussian fittings to these profiles, and summarized these position angles in the bottom panel of Figure \ref{fig:spot_fit}.
%\textcolor{cyan}{The centroid position angles of the radio emission knots which we obtained by performing Gaussian fittings to the azimuthal intensity profiles are presented in the right panel of Figure \ref{fig:jvla_intensity} (see Figure \ref{fig:spot_fit}, bottom panel).}

From both the images and the azimuthal intensity profiles we can see that during the X band observations, the intensity peak locates close to the minor axis of the DM~Tau disk (e.g., $\sim$250$^{\circ}$ position angle).
In contrast, during the third epoch of Ku band observations, the intensity peak locates close to the major axis of the DM~Tau disk (e.g., $\sim$180$^{\circ}$ position angle).
During the first epoch of Ku band observations, the intensity peak locates in between those of the former two cases ($\sim$200$^{\circ}$--250$^{\circ}$). 
Finally, from the right panel of Figure \ref{fig:jvla_intensity}, the main knot comes mainly from 10--30 AU. 
It is consistent with the expected edge of cavity of the disk, as reported by \citet{Hashimoto2021}.
Our working hypothesis to interpret these observations are given in Section \ref{sec:discussion}.

\section{Discussion} \label{sec:discussion}

\begin{figure*}
    \hspace{0.75cm}
    \begin{tabular}{cc}%{p{7.5cm} p{7.5cm}}
         \includegraphics[width=6.5cm]{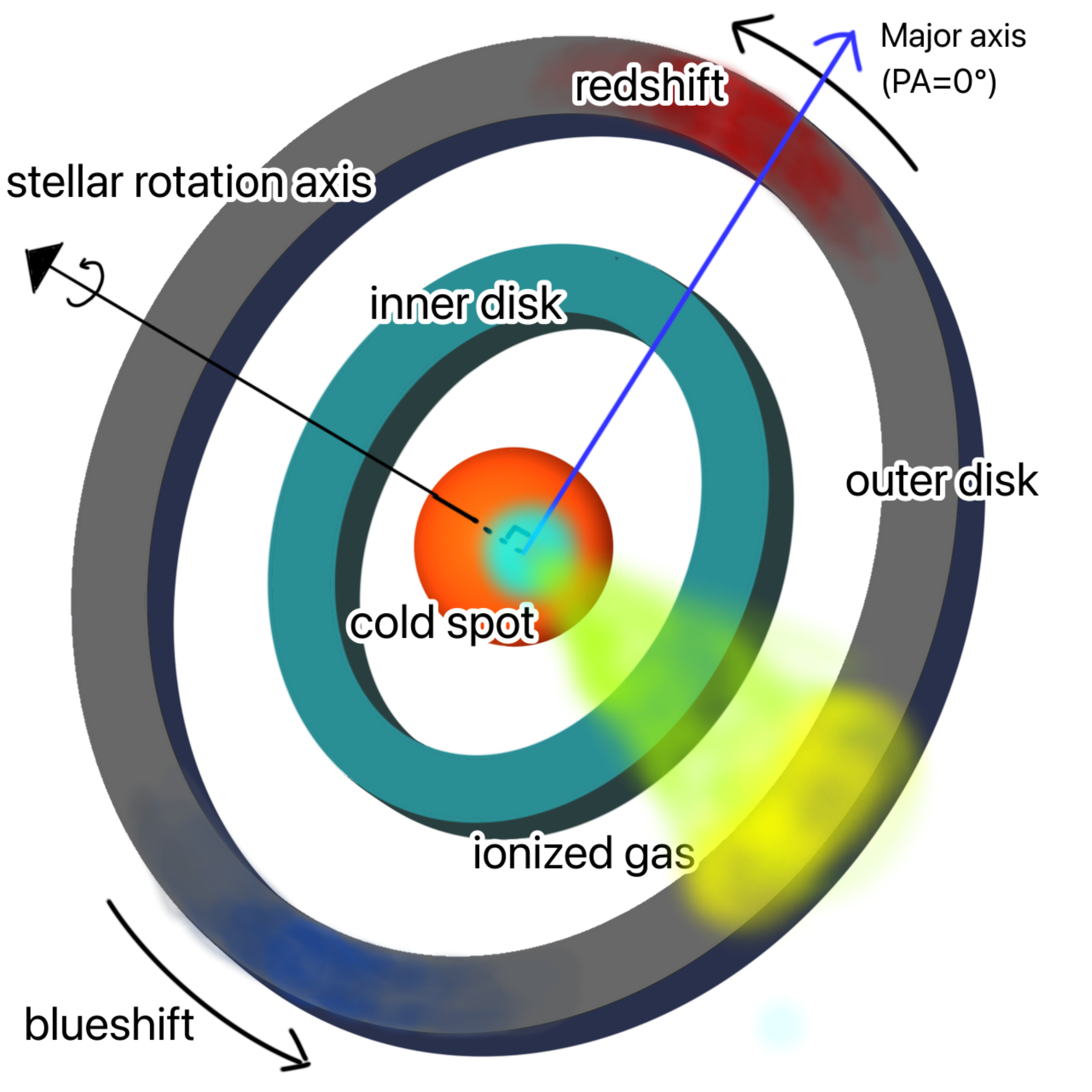} &
         \includegraphics[width=6.5cm]{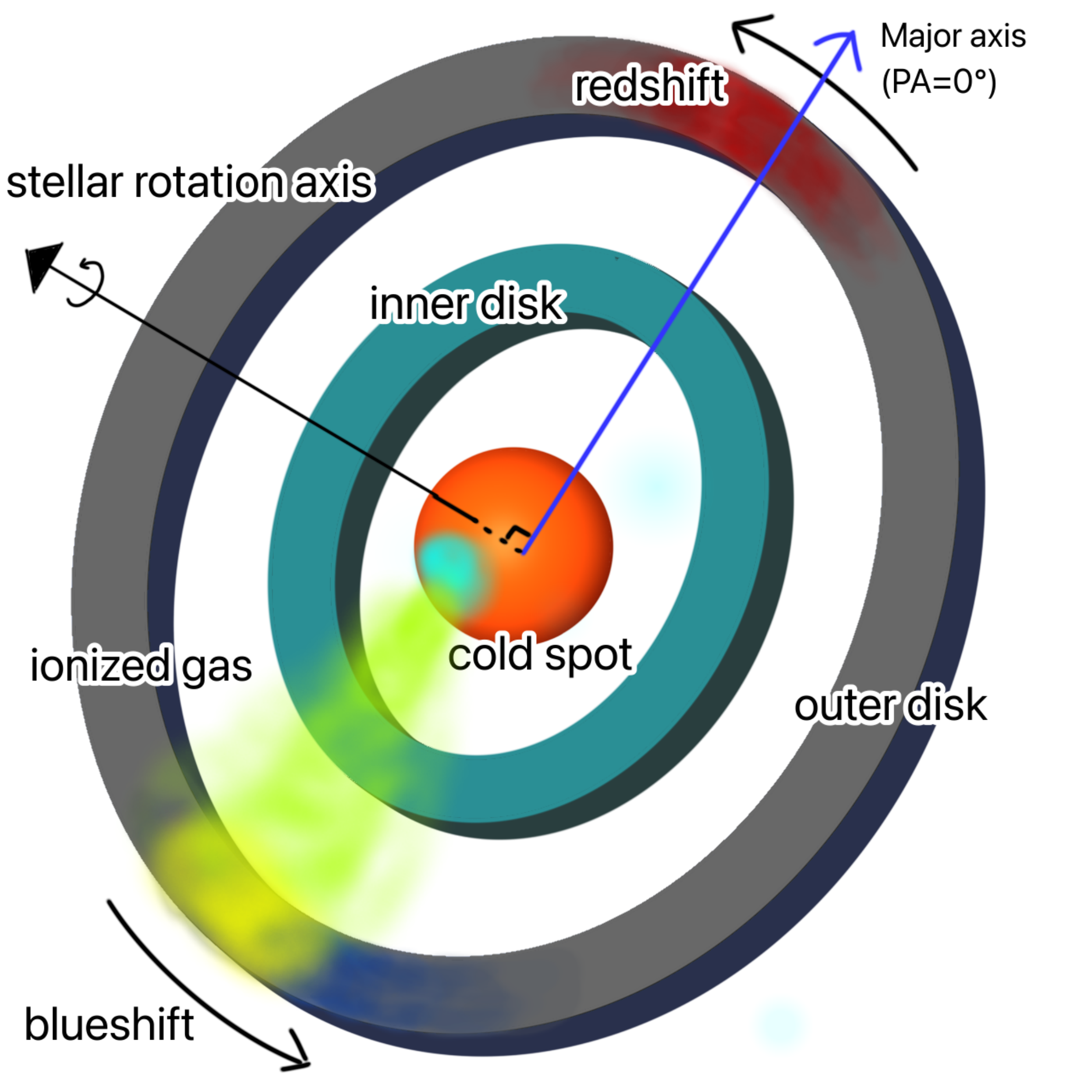} 
    \end{tabular}
    \vspace{0.5cm}
    \caption{Schematic model for interpreting the observations on DM~Tau. Left and right panels indicate the configurations during the X band (8--12 GHz) observations and the third epoch of JVLA Ku band (12--18 GHz) observations, respectively. The blue arrow indicates the major axis of the DM~Tau disk.}
    
    \vspace{0.5cm}
    \label{fig:schematic_pic}
\end{figure*}

Our working hypothesis to phenomenologically interpret the variability of the optical and infrared photometry and radio intensity distribution, is the presence of giant cold spots on DM~Tau and the anisotropic X-ray or UV illumination related to spot activities.
We conjecture that the cold spots are the dominant X-ray or UV sources that illuminate the cones which are opening from the local area of the spot (for a schematic picture, see Figure \ref{fig:schematic_pic}).
In this concept, the rotation of the host PMS star breaks the spherical symmetry of the system, making the cold spots preferentially populated close to the north and/or south poles of the overall magnetic dipole.
The spin of the host PMS stars makes the spots illuminate the ambient regions with bi-conical X-ray or UV beams like spinning lighthouses. 
The areas where the disk can be photoevaporated/ionized depend on the tilt angle of the rotation axis of the PMS star, the latitude of the spot(s), and the opening angle of the illumination cone.
Presently, the last one is the least understood.

For example, when the stellar and disk rotation axes are parallel to each other, and if the cold spots are populated at high latitudes, the photoevaporation/ionization will be minimized.
If this configuration is common, which may be expected, then the anisotropic UV or X-ray illumination may help explain why the young stars appear under-luminous in the previous centimeter surveys (e.g., \citealt{Pascucci2014ApJ...795....1P,Galvan2014A&A...570L...9G}).
In other words, the under-luminous radio emission of the PMS stars may support our present hypothesis.

As a preliminary consistency check of our interpretation, we modeled the photometric data assuming that there is a single dominant giant spot.
We employed the methodology in the \textit{SOAP} tool (\citealt{Boisse2012}).
The details are described in Appendix \ref{appendix:spotmodel}.

\begin{figure}
    \hspace{-1.5cm}
    \begin{tabular}{c}
    \includegraphics[width=9.5cm]{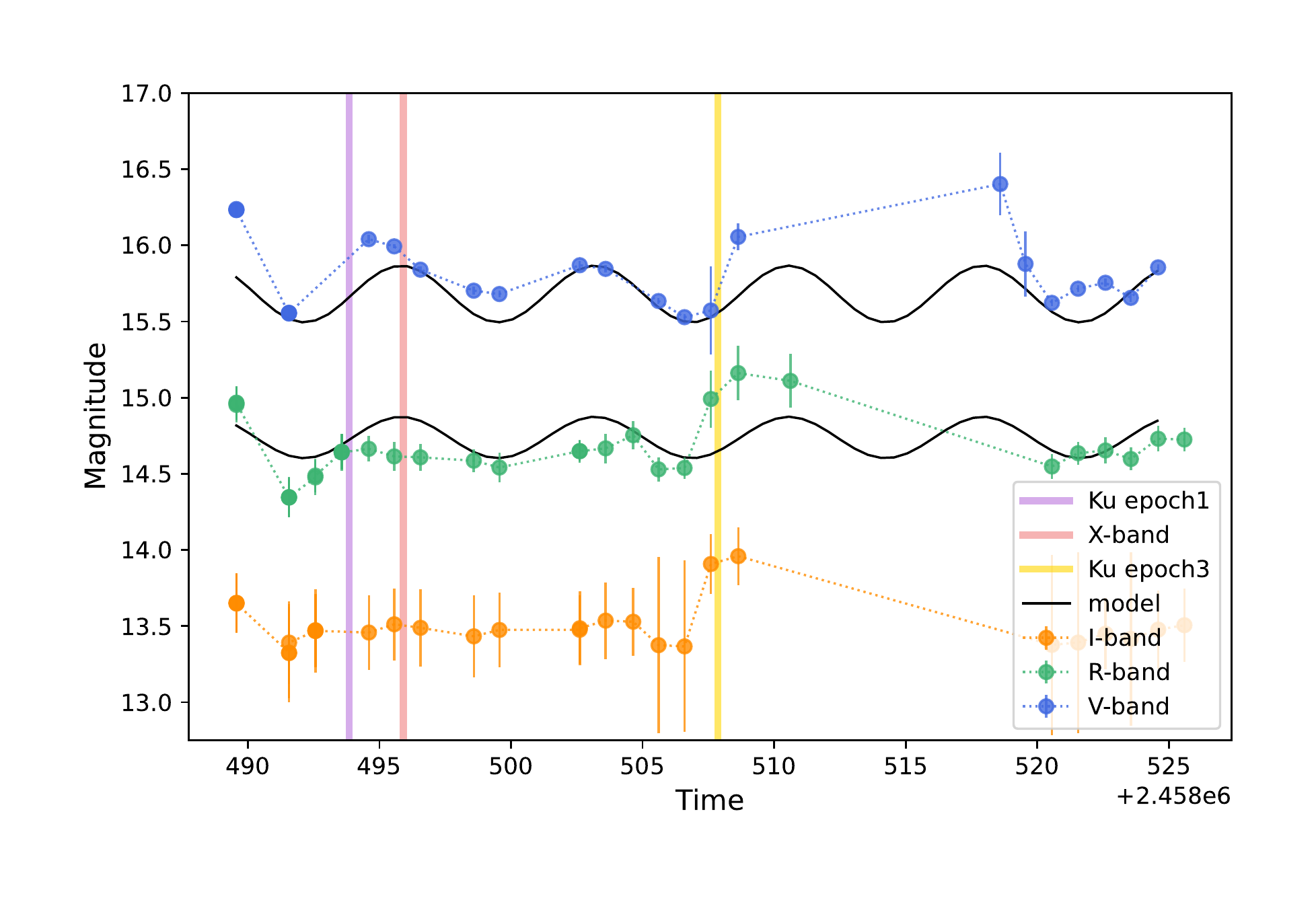} \\
    \vspace{-1.5cm}\includegraphics[width=9.5cm]{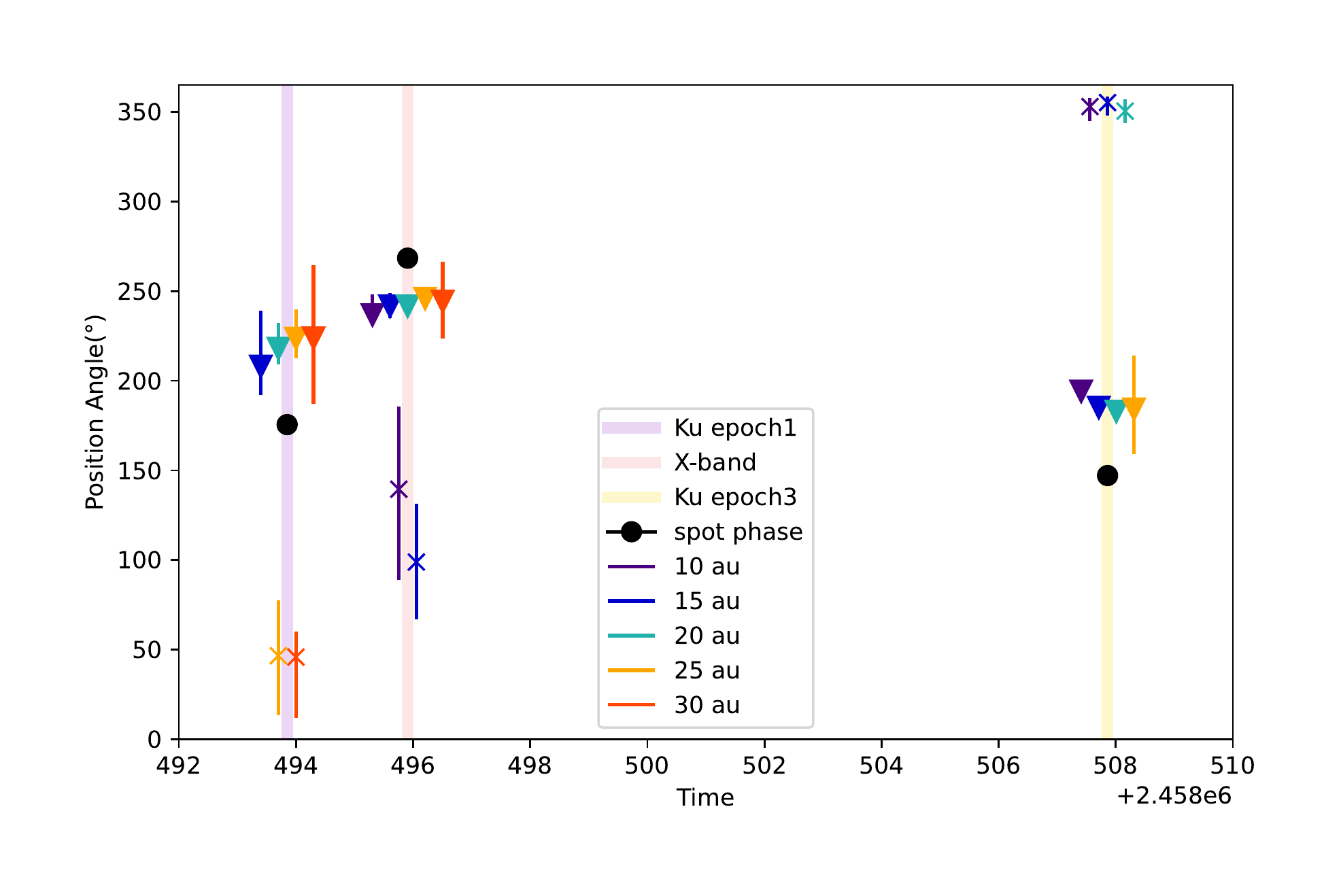}
    \end{tabular}
    \vspace{0.5cm}
    \caption{
    {\it Top: } A comparison of the magnitudes of the V, R, and I bands monitoring observations with those given by our photospheric model including a stellar spot (Appendix \ref{appendix:spotmodel}, Section \ref{sec:discussion}).
    For clarity, the magnitudes were vertically shifted in the same way as in Figure \ref{fig:monitoring}.  
    Similar to Figure \ref{fig:monitoring}, the three vertical lines indicate the (-28 periods shifted) time epochs of our successful JVLA observations.
    {\it Bottom: } The mean position angle(s) derived based on the Gaussian fittings to the azimuthal intensity profiles (Figure \ref{fig:jvla_intensity}, right). Downward triangles ($\blacktriangledown$) show the brightest peaks and cross marks show the second brightest peaks at each radius.
    The black circle indicates the spot phase (for an explanation see Section \ref{sec:discussion}).
    %\todo{The spot position angle measurements will be included in the next circulation.}
    }
    \label{fig:spot_fit}
\end{figure}

A comparison between the synthetic optical light curves produced from our model and our V and R bands observations are presented in Figure \ref{fig:spot_fit}.
We found that the periodic modulation of the V and R bands magnitudes can indeed be interpreted with a high latitude giant cold spot that is $\sim$1000 K cooler than the ambient stellar surface (Figure \ref{fig:spot_fit}, upper panel; Table \ref{tab:input_params}).
In our best-fit model, the spot covers $\sim$50\% of the stellar surface.
Both the temperature and area of our spot model are reasonable when compared to what were typically found in the M dwarfs (\citealt{Fang2016, Jackson2018, Flores2020}).
We note that the $v\sin i$ value of the host protostar derived from the previous optical spectroscopy (\citealt{Hartmann1989}) and the stellar radius we derived may imply that the host protostar is $\sim$15$^{\circ}$ more inclined than the disk.

In our simplified spot model, we assumed a single circular spot although the spot may have a more complicated shape in reality. 
It can also be aggregation of multiple spots.
To characterize how the spot is illuminating the disk without being seriously biased by the assumption of the shape of the spot(s), we define a quantity that is called {\it spot phase}.
The unit of spot phase is degree.
%We define the spot phase to be zero when the spot is phasing the southwest, and when projected onto the disk plane, is in the minor axis of the disk.
We define the spot phase to be zero when the normal vector on the spot center is pointing toward the northwest and is projectedly aligned with the major axis of the disk.
The spot phase increases counter-clockwisely by 2$\pi$ over a stellar rotational time period.
The bottom panel of Figure \ref{fig:spot_fit} shows a comparison between the spot phase and the centroids of the radio emission knots.
There appears to be a good correlation between these quantities.
We note that although our JVLA observations (Section \ref{sub:obsJVLA}) were taken $\sim$7 months after the OIR monitoring observations carried out in 2019 January (Section \ref{sub:obsBVI}).
Nevertheless, our OIR monitoring observations were carried out also about two years after the K2 observations while our monitoring observations still detected consistent periodicity and phase with those observed from the K2 data.
Therefore, we consider that it is fair to extrapolate the spot phase from the epochs of the K2 and OIR monitoring observaions to make comparison with our JVLA observations.

We hypothesize that there is one dominant giant cold spot.
During the two epochs of JVLA Ku band (12--18 GHz) observations, this cold spot center faced and illuminated the southeast of the DM~Tau disk (Figure \ref{fig:schematic_pic}, left).
Between the first epoch of Ku band observations and the X band (8--12 GHz) observations, the spot center rotated toward upper right (Figure \ref{fig:schematic_pic}, left) and thus the UV/X-ray photoionized region moved toward the minor axis of the DM~Tau disk.
The spot was approximately facing us during the X band observations and therefore the OIR magnitudes were minimized.
In this case, the sense of the protostellar rotation is consistent with that of the disk rotation traced by CO velocities (redshifted in the northwest and blueshifted in the southeast; see Figure \ref{fig:schematic_pic}; \citealt{Kudo2018}).

There might be more than one cold spot on the surface of the host protostars, as the third epoch of Ku band observations (Figure \ref{fig:jvla_intensity}, bottom) resolved more than one emission ionized emission knot.
There might be another giant spot that is located at the far-side of the stellar surface and may illuminate the northeast with UV and/or X-ray photons, although either the visual inspection of the OIR light curves or our spot modeling could not provide any evidence to either support or against this possibility.
If there is indeed a giant spot on the far-side of the stellar surface, it is still possible that its UV/X-ray photo-illumination is blocked by the inner disk (Figure \ref{fig:schematic_pic}) that might be slightly inclined with respect to the host protostar and the 19 au dust ring  (\citealt{Hashimoto2021}).
We note that the previous ALMA 1.3 mm continuum image showed that the southwest part of the 19 au ring appears brighter than the northeast part (\citealt{Hashimoto2021}).

Lastly, accretion and the associated hot spots have been suggested as a source of X-ray and/or UV radiation at classical T Tauri stars \citep{Manara2021}, and the relationship between such radiation and ionized winds has been theoretically investigated\citep{Nakatani2022}.
To check whether or not the prospective hot spots can be important X-ray or UV emission sources in DM\,Tau, we followed the definition in \citet{Grankin2007} to define the C1 and C2 parameters from our own V band photometric monitoring and then based on these parameters to assess whether the quasi-periodic OIR variability in DM\,Tau is mainly caused by hot or cold spot(s).
Since our photometric monitoring did not cover a time period that is as long as that of the ROTOR-program data presented in \citet{Grankin2007}, our derived C1 and C2 values are rather uncertain. 
Nevertheless, our derived C1 (0.508) and C2 (1.323) appear very well consistent with their reported values (C1$=$0.571, C2$=$1.285). 
Here we point out that although the C1 and C2 values are objective measurements, attributing certain C1/C2 values to cold or hot spots was an interpretation that must be based on assumptions. 
Based on certain hypotheses, \citet{Grankin2007} suggested that the C1$\sim$0.5 and C2$\sim$1 results can be interpreted by a combination of hot and cold spots. 
However, the photometric monitoring survey towards weak-line T Tauri stars (WTTS; \citealt{Grankin2008A&A...479..827G}), which are unlikely to host strong hot spots, has shown that the cold-spotted WTTS generally have such C1 ($\sim$0.5) and C2 ($\sim$1) values. Therefore, both our periodogram analysis and the C1/C2 values derived by us and \citet{Grankin2007} support that the quasi-periodic OIR variability in DM\,Tau can be caused by a cold spot or some cold spots, superimposed by the variability that might be caused by time-varying accretion.

We presently do not possess observational evidence to support a statement that the hot spots can be the ionizing photo illumination sources that lead to the observed, spatially asymmetric radio emission in DM\,Tau. 
DM Tau may also have accretion variability, while it is also not yet clear to us how the accretion variability of DM\,Tau is related to the spatially asymmetric radio emission resolved in our JVLA observations.
With only 3 epochs of radio observations, we cannot rule out the possibilities that the hot spots and accretion variability are related to the resolved radio emission by chance although we tentatively do not favor such an interpretation.

Our present hypotheses can be tested by the joint X-ray, UV, optical, and radio imaging observations for a full stellar rotation period (or longer).
Verifying these hypotheses will lead to important applications with the future high angular-resolution radio imaging facilities, such as the next-generation very large array (ngVLA) or Square Kilometer Array (SKA).

An alternative hypothesis to interpret the periodic optical modulation is that the innermost DM~Tau disk is warped and thus periodically obscures the protostellar emission from our observations.
We presently disfavor this scenario due to the following reasons.
First, its lack of infrared excess (\citealt{Calvet2005,Kudo2018}) makes it very difficult to incorporate an inner disk structure that orbits at $\sim$7 days's period.
Second, the phase of the obscuration (e.g., showing minimal OIR magnitudes during the JVLA X band observations) would make it hard to understand why the ionized gas was seen in the southeast during our X band observations.

\section{Conclusion} \label{sec:conclusion}
We have carried out OIR photometric monitoring observations and three epochs of JVLA high angular-resolution imaging observations towards the M1-type PMS star, DM~Tau. In addition, we have retrieved the K2 optical light curve on DM~Tau. Moreover, we have performed the JVLA continuum observations at X (8--12 GHz) and Ku (12--18 GHz) bands in the most extended array configuration (A array configuration).

In spite of the rich flaring activities, we found that our OIR monitoring observations and the K2 light curve show consistent periodic modulation with a 7.364 days period. It can be interpreted by a high-latitude, giant cold spot that is $\sim$1000 K cooler than the ambient stellar surface. Both the flares in the OIR light curves and the presence of a giant spot hint that DM~Tau is magnetically active.  We hypothesize that the giant cold spot may be a prominent or dominant UV or X-ray source, which is analogous to the sunspots. In addition, we conjecture that the UV and X-ray illumination is limited to a cone-like region where the location of the giant cold spot is the vertex of the cone.  With such hypothesis and conjecture, qualitatively, we can concordantly explain why our high-angular resolution JVLA observations resolved localized ionized gas that has time-varying spatial distribution. In this scenario, the host PMS star is rotating in the same sense as the natal circumstellar disk. The giant cold spot may be illuminating the ambient gas with high-energy photons like a spinning lighthouse. Anisotropy of the high-energy photo-illumination may explain why the ionized gas emission appears unexpectedly dim in the previous JVLA surveys towards nearby protostars, in particular, the spots populating at high latitudes may not illuminate the ambient protoplanetary disks efficiently. If it is indeed the case, then understanding the habitability of exoplanets and the formation of (pre-)biotic molecules would require detailed theoretical models for the formation of protostellar spots and their high-energy illumination.

%TC:ignore
\acknowledgments
This paper includes data collected by the Kepler mission. Funding for the Kepler mission is provided by the NASA Science Mission directorate.
This paper includes data collected by the K2 mission. Funding for the K2 mission is provided by the NASA Science Mission directorate.
Some/all of the data presented in this paper were obtained from the Mikulski Archive for Space Telescopes (MAST) at the Space Telescope Science Institute. The specific observations analyzed can be accessed via \dataset[10.17909/T93M48]{https://doi.org/10.17909/T93M48}.
This paper makes use of the following ALMA data: ADS/JAO.ALMA \#2018.1.01755.S, ADS/JAO.ALMA \#2017.1.01460.S, and ADS/JAO.ALMA \#2013.1.00498.S. ALMA is a partnership of ESO (representing its member states), NSF (USA) and NINS (Japan), together with NRC (Canada), MOST and ASIAA (Taiwan), and KASI (Republic of Korea), in cooperation with the Republic of Chile.
The Joint ALMA Observatory is operated by ESO, AUI NRAO and NAOJ.
This work has made use of data from the European Space Agency (ESA) mission Gaia (https://www.cosmos.esa.int/gaia), processed by the Gaia Data Processing and Analysis Consortium (DPAC, https://www.cosmos.esa.int/web/gaia/dpac/consortium). 
Funding for the DPAC has been provided by national institutions, in particular the institutions participating in the Gaia Multilateral Agreement.
This research made use of Photutils, an Astropy package for detection and photometry of astronomical sources \citet{Bradley2019}.
H.B.L. is supported by the National Science and Technology Council (NSTC) of Taiwan (Grant Nos. 111-2112-M-110-022-MY3).
This research has made use of the SVO Filter Profile Service (http://svo2.cab.inta-csic.es/theory/fps/) supported from the Spanish MINECO through grant AYA2017-84089.


%\clearpage
\vspace{2cm}
\facilities{ALMA, JVLA, Kepler-K2}


\software{astropy \citep{2013A&A...558A..33A},  
          Scipy \citep[v1.0; ][]{2019arXiv190710121V},
          Numpy \citep{VanDerWalt2011}, 
          CASA \citep[v5.6.0; ][]{McMullin2007},
          Photutils \citep{Bradley2019}
          }



\clearpage

\bibliography{sample63}{}

\begin{thebibliography}{}
\expandafter\ifx\csname natexlab\endcsname\relax\def\natexlab#1{#1}\fi
\providecommand{\url}[1]{\href{#1}{#1}}
\providecommand{\dodoi}[1]{doi:~\href{http://doi.org/#1}{\nolinkurl{#1}}}
\providecommand{\doeprint}[1]{\href{http://ascl.net/#1}{\nolinkurl{http://ascl.net/#1}}}
\providecommand{\doarXiv}[1]{\href{https://arxiv.org/abs/#1}{\nolinkurl{https://arxiv.org/abs/#1}}}

\bibitem[{{Andrews} {et~al.}(2011){Andrews}, {Wilner}, {Espaillat}, {Hughes},
  {Dullemond}, {McClure}, {Qi}, \& {Brown}}]{Andrews2011}
{Andrews}, S.~M., {Wilner}, D.~J., {Espaillat}, C., {et~al.} 2011, \apj, 732,
  42, \dodoi{10.1088/0004-637X/732/1/42}

\bibitem[{{Armitage}(2016)}]{Armitage2016ApJ...833L..15A}
{Armitage}, P.~J. 2016, \apjl, 833, L15, \dodoi{10.3847/2041-8213/833/2/L15}

\bibitem[{{Astropy Collaboration} {et~al.}(2013){Astropy Collaboration},
  {Robitaille}, {Tollerud}, {Greenfield}, {Droettboom}, {Bray}, {Aldcroft},
  {Davis}, {Ginsburg}, {Price-Whelan}, {Kerzendorf}, {Conley}, {Crighton},
  {Barbary}, {Muna}, {Ferguson}, {Grollier}, {Parikh}, {Nair}, {Unther},
  {Deil}, {Woillez}, {Conseil}, {Kramer}, {Turner}, {Singer}, {Fox}, {Weaver},
  {Zabalza}, {Edwards}, {Azalee Bostroem}, {Burke}, {Casey}, {Crawford},
  {Dencheva}, {Ely}, {Jenness}, {Labrie}, {Lim}, {Pierfederici}, {Pontzen},
  {Ptak}, {Refsdal}, {Servillat}, \& {Streicher}}]{2013A&A...558A..33A}
{Astropy Collaboration}, {Robitaille}, T.~P., {Tollerud}, E.~J., {et~al.} 2013,
  \aap, 558, A33, \dodoi{10.1051/0004-6361/201322068}

\bibitem[{{Baglin} {et~al.}(2006){Baglin}, {Auvergne}, {Barge}, {Deleuil},
  {Catala}, {Michel}, {Weiss}, \& {COROT Team}}]{Baglin2006}
{Baglin}, A., {Auvergne}, M., {Barge}, P., {et~al.} 2006, in ESA Special
  Publication, Vol. 1306, The CoRoT Mission Pre-Launch Status - Stellar
  Seismology and Planet Finding, ed. M.~{Fridlund}, A.~{Baglin}, J.~{Lochard},
  \& L.~{Conroy}, 33

\bibitem[{{Boisse} {et~al.}(2012){Boisse}, {Bonfils}, \& {Santos}}]{Boisse2012}
{Boisse}, I., {Bonfils}, X., \& {Santos}, N.~C. 2012, \aap, 545, A109,
  \dodoi{10.1051/0004-6361/201219115}

\bibitem[{Bradley {et~al.}(2019)Bradley, Sip{\H o}cz, Robitaille, Tollerud,
  Vin{\'{\i}}cius, Deil, Barbary, G{\"u}nther, Cara, Busko, Conseil,
  Droettboom, Bostroem, Bray, Bratholm, Wilson, Craig, Barentsen, Pascual,
  Donath, Greco, Perren, Lim, \& Kerzendorf}]{Bradley2019}
Bradley, L., Sip{\H o}cz, B., Robitaille, T., {et~al.} 2019, astropy/photutils:
  v0.6, \dodoi{10.5281/zenodo.2533376}

\bibitem[{{Calvet} {et~al.}(2005){Calvet}, {D'Alessio}, {Watson},
  {Franco-Hern{\'a}ndez}, {Furlan}, {Green}, {Sutter}, {Forrest}, {Hartmann},
  {Uchida}, {Keller}, {Sargent}, {Najita}, {Herter}, {Barry}, \&
  {Hall}}]{Calvet2005}
{Calvet}, N., {D'Alessio}, P., {Watson}, D.~M., {et~al.} 2005, \apjl, 630,
  L185, \dodoi{10.1086/491652}

\bibitem[{{Chang} {et~al.}(2018){Chang}, {Lin}, {Ip}, {Huang}, {Hou}, {Yu},
  {Song}, \& {Luo}}]{Chang2018ApJ...867...78C}
{Chang}, H.~Y., {Lin}, C.~L., {Ip}, W.~H., {et~al.} 2018, \apj, 867, 78,
  \dodoi{10.3847/1538-4357/aae2bc}

\bibitem[{{Claret} {et~al.}(2012){Claret}, {Hauschildt}, \&
  {Witte}}]{Claret2012}
{Claret}, A., {Hauschildt}, P.~H., \& {Witte}, S. 2012, \aap, 546, A14,
  \dodoi{10.1051/0004-6361/201219849}

\bibitem[{{Cleeves} {et~al.}(2017){Cleeves}, {Bergin}, {{\"O}berg}, {Andrews},
  {Wilner}, \& {Loomis}}]{Cleeves2017}
{Cleeves}, L.~I., {Bergin}, E.~A., {{\"O}berg}, K.~I., {et~al.} 2017, \apjl,
  843, L3, \dodoi{10.3847/2041-8213/aa76e2}

\bibitem[{{Cody} \& {Hillenbrand}(2018)}]{Cody2018}
{Cody}, A.~M., \& {Hillenbrand}, L.~A. 2018, \aj, 156, 71,
  \dodoi{10.3847/1538-3881/aacead}

\bibitem[{{Cody} {et~al.}(2022){Cody}, {Hillenbrand}, \& {Rebull}}]{Cody2022}
{Cody}, A.~M., {Hillenbrand}, L.~A., \& {Rebull}, L.~M. 2022, \aj, 163, 212,
  \dodoi{10.3847/1538-3881/ac5b73}

\bibitem[{{Cody} {et~al.}(2014){Cody}, {Stauffer}, {Baglin}, {Micela},
  {Rebull}, {Flaccomio}, {Morales-Calder{\'o}n}, {Aigrain}, {Bouvier},
  {Hillenbrand}, {Gutermuth}, {Song}, {Turner}, {Alencar}, {Zwintz},
  {Plavchan}, {Carpenter}, {Findeisen}, {Carey}, {Terebey}, {Hartmann},
  {Calvet}, {Teixeira}, {Vrba}, {Wolk}, {Covey}, {Poppenhaeger}, {G{\"u}nther},
  {Forbrich}, {Whitney}, {Affer}, {Herbst}, {Hora}, {Barrado}, {Holtzman},
  {Marchis}, {Wood}, {Medeiros Guimar{\~a}es}, {Lillo Box}, {Gillen},
  {McQuillan}, {Espaillat}, {Allen}, {D'Alessio}, \& {Favata}}]{Cody2014}
{Cody}, A.~M., {Stauffer}, J., {Baglin}, A., {et~al.} 2014, \aj, 147, 82,
  \dodoi{10.1088/0004-6256/147/4/82}

\bibitem[{{Cornwell}(2008)}]{Cornwell2008MultiscaleClean}
{Cornwell}, T.~J. 2008, IEEE Journal of Selected Topics in Signal Processing,
  2, 793, \dodoi{10.1109/JSTSP.2008.2006388}

\bibitem[{{Coutens} {et~al.}(2019){Coutens}, {Liu}, {Jim{\'e}nez-Serra},
  {Bourke}, {Forbrich}, {Hoare}, {Loinard}, {Testi}, {Audard}, {Caselli},
  {Chac{\'o}n-Tanarro}, {Codella}, {Di Francesco}, {Fontani}, {Hogerheijde},
  {Johansen}, {Johnstone}, {Maddison}, {Pani{\'c}}, {P{\'e}rez}, {Podio},
  {Punanova}, {Rawlings}, {Semenov}, {Tazzari}, {Tobin}, {van der Wiel}, {van
  Langevelde}, {Vlemmings}, {Walsh}, \& {Wilner}}]{Coutens2019A&A...631A..58C}
{Coutens}, A., {Liu}, H.~B., {Jim{\'e}nez-Serra}, I., {et~al.} 2019, \aap, 631,
  A58, \dodoi{10.1051/0004-6361/201935340}

\bibitem[{{Dong} \& {Fung}(2017)}]{Dong2017ApJ...835..146D}
{Dong}, R., \& {Fung}, J. 2017, \apj, 835, 146,
  \dodoi{10.3847/1538-4357/835/2/146}

\bibitem[{{Dzib} {et~al.}(2015){Dzib}, {Loinard}, {Rodr{\'\i}guez},
  {Mioduszewski}, {Ortiz-Le{\'o}n}, {Kounkel}, {Pech}, {Rivera}, {Torres},
  {Boden}, {Hartmann}, {Evans}, {Brice{\~n}o}, \&
  {Tobin}}]{Dzib2015ApJ...801...91D}
{Dzib}, S.~A., {Loinard}, L., {Rodr{\'\i}guez}, L.~F., {et~al.} 2015, \apj,
  801, 91, \dodoi{10.1088/0004-637X/801/2/91}

\bibitem[{{Edwards} {et~al.}(1993){Edwards}, {Strom}, {Hartigan}, {Strom},
  {Hillenbrand}, {Herbst}, {Attridge}, {Merrill}, {Probst}, \&
  {Gatley}}]{Edwards1993}
{Edwards}, S., {Strom}, S.~E., {Hartigan}, P., {et~al.} 1993, \aj, 106, 372,
  \dodoi{10.1086/116646}

\bibitem[{{Espaillat} {et~al.}(2022){Espaillat}, {Mac{\'\i}as}, {Wendeborn},
  {Franco-Hern{\'a}ndez}, {Calvet}, {Rilinger}, {Cleeves}, \&
  {D'Alessio}}]{Espaillat2022ApJ...924..104E}
{Espaillat}, C.~C., {Mac{\'\i}as}, E., {Wendeborn}, J., {et~al.} 2022, \apj,
  924, 104, \dodoi{10.3847/1538-4357/ac365a}

\bibitem[{{Fang} {et~al.}(2016){Fang}, {Zhao}, {Zhao}, {Chen}, \& {Bharat
  Kumar}}]{Fang2016}
{Fang}, X.-S., {Zhao}, G., {Zhao}, J.-K., {Chen}, Y.-Q., \& {Bharat Kumar}, Y.
  2016, \mnras, 463, 2494, \dodoi{10.1093/mnras/stw1923}

\bibitem[{{Favata} {et~al.}(2005){Favata}, {Flaccomio}, {Reale}, {Micela},
  {Sciortino}, {Shang}, {Stassun}, \& {Feigelson}}]{Favata2005}
{Favata}, F., {Flaccomio}, E., {Reale}, F., {et~al.} 2005, \apjs, 160, 469,
  \dodoi{10.1086/432542}

\bibitem[{{Feigelson} \& {Montmerle}(1999)}]{Feigelson1999}
{Feigelson}, E.~D., \& {Montmerle}, T. 1999, \araa, 37, 363,
  \dodoi{10.1146/annurev.astro.37.1.363}

\bibitem[{{Flores} {et~al.}(2020){Flores}, {Reipurth}, \&
  {Connelley}}]{Flores2020}
{Flores}, C., {Reipurth}, B., \& {Connelley}, M.~S. 2020, \apj, 898, 109,
  \dodoi{10.3847/1538-4357/ab9e67}

\bibitem[{{Gaia Collaboration} {et~al.}(2021){Gaia Collaboration}, {Smart},
  {Sarro}, {Rybizki}, {Reyl{\'e}}, {Robin}, {Hambly}, {Abbas}, {Barstow}, {de
  Bruijne}, {Bucciarelli}, {Carrasco}, {Cooper}, {Hodgkin}, {Masana},
  {Michalik}, {Sahlmann}, {Sozzetti}, {Brown}, {Vallenari}, {Prusti},
  {Babusiaux}, {Biermann}, {Creevey}, {Evans}, {Eyer}, {Hutton}, {Jansen},
  {Jordi}, {Klioner}, {Lammers}, {Lindegren}, {Luri}, {Mignard}, {Panem},
  {Pourbaix}, {Randich}, {Sartoretti}, {Soubiran}, {Walton}, {Arenou},
  {Bailer-Jones}, {Bastian}, {Cropper}, {Drimmel}, {Katz}, {Lattanzi}, {van
  Leeuwen}, {Bakker}, {Casta{\~n}eda}, {De Angeli}, {Ducourant}, {Fabricius},
  {Fouesneau}, {Fr{\'e}mat}, {Guerra}, {Guerrier}, {Guiraud}, {Jean-Antoine
  Piccolo}, {Messineo}, {Mowlavi}, {Nicolas}, {Nienartowicz}, {Pailler},
  {Panuzzo}, {Riclet}, {Roux}, {Seabroke}, {Sordo}, {Tanga}, {Th{\'e}venin},
  {Gracia-Abril}, {Portell}, {Teyssier}, {Altmann}, {Andrae}, {Bellas-Velidis},
  {Benson}, {Berthier}, {Blomme}, {Brugaletta}, {Burgess}, {Busso}, {Carry},
  {Cellino}, {Cheek}, {Clementini}, {Damerdji}, {Davidson}, {Delchambre},
  {Dell'Oro}, {Fern{\'a}ndez-Hern{\'a}ndez}, {Galluccio}, {Garc{\'\i}a-Lario},
  {Garcia-Reinaldos}, {Gonz{\'a}lez-N{\'u}{\~n}ez}, {Gosset}, {Haigron},
  {Halbwachs}, {Harrison}, {Hatzidimitriou}, {Heiter}, {Hern{\'a}ndez},
  {Hestroffer}, {Holl}, {Jan{\ss}en}, {Jevardat de Fombelle}, {Jordan},
  {Krone-Martins}, {Lanzafame}, {L{\"o}ffler}, {Lorca}, {Manteiga}, {Marchal},
  {Marrese}, {Moitinho}, {Mora}, {Muinonen}, {Osborne}, {Pancino}, {Pauwels},
  {Recio-Blanco}, {Richards}, {Riello}, {Rimoldini}, {Roegiers}, {Siopis},
  {Smith}, {Ulla}, {Utrilla}, {van Leeuwen}, {van Reeven}, {Abreu Aramburu},
  {Accart}, {Aerts}, {Aguado}, {Ajaj}, {Altavilla}, {{\'A}lvarez}, {{\'A}lvarez
  Cid-Fuentes}, {Alves}, {Anderson}, {Anglada Varela}, {Antoja}, {Audard},
  {Baines}, {Baker}, {Balaguer-N{\'u}{\~n}ez}, {Balbinot}, {Balog}, {Barache},
  {Barbato}, {Barros}, {Bartolom{\'e}}, {Bassilana}, {Bauchet},
  {Baudesson-Stella}, {Becciani}, {Bellazzini}, {Bernet}, {Bertone}, {Bianchi},
  {Blanco-Cuaresma}, {Boch}, {Bombrun}, {Bossini}, {Bouquillon}, {Bragaglia},
  {Bramante}, {Breedt}, {Bressan}, {Brouillet}, {Burlacu}, {Busonero},
  {Butkevich}, {Buzzi}, {Caffau}, {Cancelliere}, {C{\'a}novas},
  {Cantat-Gaudin}, {Carballo}, {Carlucci}, {Carnerero}, {Casamiquela},
  {Castellani}, {Castro-Ginard}, {Castro Sampol}, {Chaoul}, {Charlot},
  {Chemin}, {Chiavassa}, {Cioni}, {Comoretto}, {Cornez}, {Cowell}, {Crifo},
  {Crosta}, {Crowley}, {Dafonte}, {Dapergolas}, {David}, {David}, {de Laverny},
  {De Luise}, {De March}, {De Ridder}, {de Souza}, {de Teodoro}, {de Torres},
  {del Peloso}, {del Pozo}, {Delgado}, {Delgado}, {Delisle}, {Di Matteo},
  {Diakite}, {Diener}, {Distefano}, {Dolding}, {Eappachen}, {Edvardsson},
  {Enke}, {Esquej}, {Fabre}, {Fabrizio}, {Faigler}, {Fedorets}, {Fernique},
  {Fienga}, {Figueras}, {Fouron}, {Fragkoudi}, {Fraile}, {Franke}, {Gai},
  {Garabato}, {Garcia-Gutierrez}, {Garc{\'\i}a-Torres}, {Garofalo}, {Gavras},
  {Gerlach}, {Geyer}, {Giacobbe}, {Gilmore}, {Girona}, {Giuffrida}, {Gomel},
  {Gomez}, {Gonzalez-Santamaria}, {Gonz{\'a}lez-Vidal}, {Granvik},
  {Guti{\'e}rrez-S{\'a}nchez}, {Guy}, {Hauser}, {Haywood}, {Helmi}, {Hidalgo},
  {Hilger}, {H{\l}adczuk}, {Hobbs}, {Holland}, {Huckle}, {Jasniewicz},
  {Jonker}, {Juaristi Campillo}, {Julbe}, {Karbevska}, {Kervella}, {Khanna},
  {Kochoska}, {Kontizas}, {Kordopatis}, {Korn}, {Kostrzewa-Rutkowska},
  {Kruszy{\'n}ska}, {Lambert}, {Lanza}, {Lasne}, {Le Campion}, {Le Fustec},
  {Lebreton}, {Lebzelter}, {Leccia}, {Leclerc}, {Lecoeur-Taibi}, {Liao},
  {Licata}, {Lindstr{\o}m}, {Lister}, {Livanou}, {Lobel}, {Madrero Pardo},
  {Managau}, {Mann}, {Marchant}, {Marconi}, {Marcos Santos}, {Marinoni},
  {Marocco}, {Marshall}, {Martin Polo}, {Mart{\'\i}n-Fleitas}, {Masip},
  {Massari}, {Mastrobuono-Battisti}, {Mazeh}, {McMillan}, {Messina}, {Millar},
  {Mints}, {Molina}, {Molinaro}, {Moln{\'a}r}, {Montegriffo}, {Mor},
  {Morbidelli}, {Morel}, {Morris}, {Mulone}, {Munoz}, {Muraveva}, {Murphy},
  {Musella}, {Noval}, {Ord{\'e}novic}, {Orr{\`u}}, {Osinde}, {Pagani},
  {Pagano}, {Palaversa}, {Palicio}, {Panahi}, {Pawlak}, {Pe{\~n}alosa
  Esteller}, {Penttil{\"a}}, {Piersimoni}, {Pineau}, {Plachy}, {Plum},
  {Poggio}, {Poretti}, {Poujoulet}, {Pr{\v{s}}a}, {Pulone}, {Racero},
  {Ragaini}, {Rainer}, {Raiteri}, {Rambaux}, {Ramos}, {Ramos-Lerate}, {Re
  Fiorentin}, {Regibo}, {Ripepi}, {Riva}, {Rixon}, {Robichon}, {Robin},
  {Roelens}, {Rohrbasser}, {Romero-G{\'o}mez}, {Rowell}, {Royer}, {Rybicki},
  {Sadowski}, {Sagrist{\`a} Sell{\'e}s}, {Salgado}, {Salguero}, {Samaras},
  {Sanchez Gimenez}, {Sanna}, {Santove{\~n}a}, {Sarasso}, {Schultheis},
  {Sciacca}, {Segol}, {Segovia}, {S{\'e}gransan}, {Semeux}, {Shahaf},
  {Siddiqui}, {Siebert}, {Siltala}, {Slezak}, {Solano}, {Solitro}, {Souami},
  {Souchay}, {Spagna}, {Spoto}, {Steele}, {Steidelm{\"u}ller}, {Stephenson},
  {S{\"u}veges}, {Szabados}, {Szegedi-Elek}, {Taris}, {Tauran}, {Taylor},
  {Teixeira}, {Thuillot}, {Tonello}, {Torra}, {Torra}, {Turon}, {Unger},
  {Vaillant}, {van Dillen}, {Vanel}, {Vecchiato}, {Viala}, {Vicente},
  {Voutsinas}, {Weiler}, {Wevers}, {Wyrzykowski}, {Yoldas}, {Yvard}, {Zhao},
  {Zorec}, {Zucker}, {Zurbach}, \& {Zwitter}}]{gaia2020}
{Gaia Collaboration}, {Smart}, R.~L., {Sarro}, L.~M., {et~al.} 2021, \aap, 649,
  A6, \dodoi{10.1051/0004-6361/202039498}

\bibitem[{{Galv{\'a}n-Madrid} {et~al.}(2014){Galv{\'a}n-Madrid}, {Liu},
  {Manara}, {Forbrich}, {Pascucci}, {Carrasco-Gonz{\'a}lez}, {Goddi},
  {Hasegawa}, {Takami}, \& {Testi}}]{Galvan2014A&A...570L...9G}
{Galv{\'a}n-Madrid}, R., {Liu}, H.~B., {Manara}, C.~F., {et~al.} 2014, \aap,
  570, L9, \dodoi{10.1051/0004-6361/201424630}

\bibitem[{{Grankin} {et~al.}(2008){Grankin}, {Bouvier}, {Herbst}, \&
  {Melnikov}}]{Grankin2008A&A...479..827G}
{Grankin}, K.~N., {Bouvier}, J., {Herbst}, W., \& {Melnikov}, S.~Y. 2008, \aap,
  479, 827, \dodoi{10.1051/0004-6361:20078476}

\bibitem[{{Grankin} {et~al.}(2007){Grankin}, {Melnikov}, {Bouvier}, {Herbst},
  \& {Shevchenko}}]{Grankin2007}
{Grankin}, K.~N., {Melnikov}, S.~Y., {Bouvier}, J., {Herbst}, W., \&
  {Shevchenko}, V.~S. 2007, \aap, 461, 183, \dodoi{10.1051/0004-6361:20065489}

\bibitem[{{Hartmann} \& {Stauffer}(1989)}]{Hartmann1989}
{Hartmann}, L., \& {Stauffer}, J.~R. 1989, \aj, 97, 873, \dodoi{10.1086/115033}

\bibitem[{{Hashimoto} {et~al.}(2021){Hashimoto}, {Muto}, {Dong}, {Liu}, {van
  der Marel}, {Francis}, {Hasegawa}, \& {Tsukagoshi}}]{Hashimoto2021}
{Hashimoto}, J., {Muto}, T., {Dong}, R., {et~al.} 2021, \apj, 911, 5,
  \dodoi{10.3847/1538-4357/abe59f}

\bibitem[{{Hawley} {et~al.}(2014){Hawley}, {Davenport}, {Kowalski},
  {Wisniewski}, {Hebb}, {Deitrick}, \& {Hilton}}]{Hawley2014ApJ...797..121H}
{Hawley}, S.~L., {Davenport}, J. R.~A., {Kowalski}, A.~F., {et~al.} 2014, \apj,
  797, 121, \dodoi{10.1088/0004-637X/797/2/121}

\bibitem[{{Henning} {et~al.}(2010){Henning}, {Semenov}, {Guilloteau}, {Dutrey},
  {Hersant}, {Wakelam}, {Chapillon}, {Launhardt}, {Pi{\'e}tu}, \&
  {Schreyer}}]{Henning2010}
{Henning}, T., {Semenov}, D., {Guilloteau}, S., {et~al.} 2010, \apj, 714, 1511,
  \dodoi{10.1088/0004-637X/714/2/1511}

\bibitem[{{Howell} {et~al.}(2014){Howell}, {Sobeck}, {Haas}, {Still},
  {Barclay}, {Mullally}, {Troeltzsch}, {Aigrain}, {Bryson}, {Caldwell},
  {Chaplin}, {Cochran}, {Huber}, {Marcy}, {Miglio}, {Najita}, {Smith},
  {Twicken}, \& {Fortney}}]{Howell2014}
{Howell}, S.~B., {Sobeck}, C., {Haas}, M., {et~al.} 2014, \pasp, 126, 398,
  \dodoi{10.1086/676406}

\bibitem[{{Jackson} {et~al.}(2018){Jackson}, {Deliyannis}, \&
  {Jeffries}}]{Jackson2018}
{Jackson}, R.~J., {Deliyannis}, C.~P., \& {Jeffries}, R.~D. 2018, \mnras, 476,
  3245, \dodoi{10.1093/mnras/sty374}

\bibitem[{{Kanagawa} {et~al.}(2015){Kanagawa}, {Muto}, {Tanaka}, {Tanigawa},
  {Takeuchi}, {Tsukagoshi}, \& {Momose}}]{Kanagawa2015}
{Kanagawa}, K.~D., {Muto}, T., {Tanaka}, H., {et~al.} 2015, \apjl, 806, L15,
  \dodoi{10.1088/2041-8205/806/1/L15}

\bibitem[{{K{\'o}sp{\'a}l} {et~al.}(2018){K{\'o}sp{\'a}l}, {{\'A}brah{\'a}m},
  {Zsidi}, {Vida}, {Szab{\'o}}, {Mo{\'o}r}, \& {P{\'a}l}}]{Kospal2018}
{K{\'o}sp{\'a}l}, {\'A}., {{\'A}brah{\'a}m}, P., {Zsidi}, G., {et~al.} 2018,
  \apj, 862, 44, \dodoi{10.3847/1538-4357/aacafa}

\bibitem[{{Kudo} {et~al.}(2018){Kudo}, {Hashimoto}, {Muto}, {Liu}, {Dong},
  {Hasegawa}, {Tsukagoshi}, \& {Konishi}}]{Kudo2018}
{Kudo}, T., {Hashimoto}, J., {Muto}, T., {et~al.} 2018, \apjl, 868, L5,
  \dodoi{10.3847/2041-8213/aaeb1c}

\bibitem[{{Lastennet} {et~al.}(1999){Lastennet}, {Valls-Gabaud}, {Lejeune}, \&
  {Oblak}}]{Lastennet1999}
{Lastennet}, E., {Valls-Gabaud}, D., {Lejeune}, T., \& {Oblak}, E. 1999, arXiv
  e-prints, astro.
\newblock \doarXiv{astro-ph/9905334}

\bibitem[{{Liu} {et~al.}(2014){Liu}, {Galv{\'a}n-Madrid}, {Forbrich},
  {Rodr{\'\i}guez}, {Takami}, {Costigan}, {Manara}, {Yan}, {Karr}, {Chou},
  {Ho}, \& {Zhang}}]{Liu2014ApJ...780..155L}
{Liu}, H.~B., {Galv{\'a}n-Madrid}, R., {Forbrich}, J., {et~al.} 2014, \apj,
  780, 155, \dodoi{10.1088/0004-637X/780/2/155}

\bibitem[{{Manara} {et~al.}(2021){Manara}, {Frasca}, {Venuti}, {Siwak},
  {Herczeg}, {Calvet}, {Hernandez}, {Tychoniec}, {Gangi}, {Alcal{\'a}},
  {Boffin}, {Nisini}, {Robberto}, {Briceno}, {Campbell-White},
  {Sicilia-Aguilar}, {McGinnis}, {Fedele}, {K{\'o}sp{\'a}l}, {{\'A}brah{\'a}m},
  {Alonso-Santiago}, {Antoniucci}, {Arulanantham}, {Bacciotti}, {Banzatti},
  {Beccari}, {Benisty}, {Biazzo}, {Bouvier}, {Cabrit}, {Caratti o Garatti},
  {Coffey}, {Covino}, {Dougados}, {Eisl{\"o}ffel}, {Ercolano}, {Espaillat},
  {Erkal}, {Facchini}, {Fang}, {Fiorellino}, {Fischer}, {France}, {Gameiro},
  {Garcia Lopez}, {Giannini}, {Ginski}, {Grankin}, {G{\"u}nther}, {Hartmann},
  {Hillenbrand}, {Hussain}, {James}, {Koutoulaki}, {Lodato}, {Mauc{\'o}},
  {Mendigut{\'\i}a}, {Mentel}, {Miotello}, {Oudmaijer}, {Rigliaco}, {Rosotti},
  {Sanchis}, {Schneider}, {Spina}, {Stelzer}, {Testi}, {Thanathibodee}, {Vink},
  {Walter}, {Williams}, \& {Zsidi}}]{Manara2021}
{Manara}, C.~F., {Frasca}, A., {Venuti}, L., {et~al.} 2021, \aap, 650, A196,
  \dodoi{10.1051/0004-6361/202140639}

\bibitem[{{Masuda} \& {Winn}(2020)}]{Masuda2020}
{Masuda}, K., \& {Winn}, J.~N. 2020, \aj, 159, 81,
  \dodoi{10.3847/1538-3881/ab65be}

\bibitem[{{McMullin} {et~al.}(2007){McMullin}, {Waters}, {Schiebel}, {Young},
  \& {Golap}}]{McMullin2007}
{McMullin}, J.~P., {Waters}, B., {Schiebel}, D., {Young}, W., \& {Golap}, K.
  2007, in Astronomical Society of the Pacific Conference Series, Vol. 376,
  Astronomical Data Analysis Software and Systems XVI, ed. R.~A. {Shaw},
  F.~{Hill}, \& D.~J. {Bell}, 127

\bibitem[{{Monsch} {et~al.}(2019){Monsch}, {Ercolano}, {Picogna}, {Preibisch},
  \& {Rau}}]{Monsch2019}
{Monsch}, K., {Ercolano}, B., {Picogna}, G., {Preibisch}, T., \& {Rau}, M.~M.
  2019, \mnras, 483, 3448, \dodoi{10.1093/mnras/sty3346}

\bibitem[{{Nakatani} \& {Takasao}(2022)}]{Nakatani2022}
{Nakatani}, R., \& {Takasao}, S. 2022, \apj, 930, 124,
  \dodoi{10.3847/1538-4357/ac63a0}

\bibitem[{{Owen} {et~al.}(2010){Owen}, {Ercolano}, {Clarke}, \&
  {Alexander}}]{Owen2010}
{Owen}, J.~E., {Ercolano}, B., {Clarke}, C.~J., \& {Alexander}, R.~D. 2010,
  \mnras, 401, 1415, \dodoi{10.1111/j.1365-2966.2009.15771.x}

\bibitem[{{Owen} {et~al.}(2013){Owen}, {Scaife}, \& {Ercolano}}]{Owen2013}
{Owen}, J.~E., {Scaife}, A. M.~M., \& {Ercolano}, B. 2013, \mnras, 434, 3378,
  \dodoi{10.1093/mnras/stt1254}

\bibitem[{{Pascucci} {et~al.}(2012){Pascucci}, {Gorti}, \&
  {Hollenbach}}]{Pascucci2012}
{Pascucci}, I., {Gorti}, U., \& {Hollenbach}, D. 2012, \apjl, 751, L42,
  \dodoi{10.1088/2041-8205/751/2/L42}

\bibitem[{{Pascucci} {et~al.}(2014){Pascucci}, {Ricci}, {Gorti}, {Hollenbach},
  {Hendler}, {Brooks}, \& {Contreras}}]{Pascucci2014ApJ...795....1P}
{Pascucci}, I., {Ricci}, L., {Gorti}, U., {et~al.} 2014, \apj, 795, 1,
  \dodoi{10.1088/0004-637X/795/1/1}

\bibitem[{{Perley} \& {Butler}(2017)}]{Perley2017}
{Perley}, R.~A., \& {Butler}, B.~J. 2017, \apjs, 230, 7,
  \dodoi{10.3847/1538-4365/aa6df9}

\bibitem[{{Rau} \& {Cornwell}(2011)}]{Rau2011MultiscaleClean}
{Rau}, U., \& {Cornwell}, T.~J. 2011, \aap, 532, A71,
  \dodoi{10.1051/0004-6361/201117104}

\bibitem[{{Rebull} {et~al.}(2020){Rebull}, {Stauffer}, {Cody}, {Hillenbrand},
  {Bouvier}, {Roggero}, \& {David}}]{Rebull2020}
{Rebull}, L.~M., {Stauffer}, J.~R., {Cody}, A.~M., {et~al.} 2020, \aj, 159,
  273, \dodoi{10.3847/1538-3881/ab893c}

\bibitem[{{Ricker} {et~al.}(2015){Ricker}, {Winn}, {Vanderspek}, {Latham},
  {Bakos}, {Bean}, {Berta-Thompson}, {Brown}, {Buchhave}, {Butler}, {Butler},
  {Chaplin}, {Charbonneau}, {Christensen-Dalsgaard}, {Clampin}, {Deming},
  {Doty}, {De Lee}, {Dressing}, {Dunham}, {Endl}, {Fressin}, {Ge}, {Henning},
  {Holman}, {Howard}, {Ida}, {Jenkins}, {Jernigan}, {Johnson}, {Kaltenegger},
  {Kawai}, {Kjeldsen}, {Laughlin}, {Levine}, {Lin}, {Lissauer}, {MacQueen},
  {Marcy}, {McCullough}, {Morton}, {Narita}, {Paegert}, {Palle}, {Pepe},
  {Pepper}, {Quirrenbach}, {Rinehart}, {Sasselov}, {Sato}, {Seager},
  {Sozzetti}, {Stassun}, {Sullivan}, {Szentgyorgyi}, {Torres}, {Udry}, \&
  {Villasenor}}]{Ricker2015}
{Ricker}, G.~R., {Winn}, J.~N., {Vanderspek}, R., {et~al.} 2015, Journal of
  Astronomical Telescopes, Instruments, and Systems, 1, 014003,
  \dodoi{10.1117/1.JATIS.1.1.014003}

\bibitem[{{Rodrigo} \& {Solano}(2020)}]{Rodrigo2020}
{Rodrigo}, C., \& {Solano}, E. 2020, in XIV.0 Scientific Meeting (virtual) of
  the Spanish Astronomical Society, 182

\bibitem[{{Rodrigo} {et~al.}(2012){Rodrigo}, {Solano}, \& {Bayo}}]{Rodrigo2012}
{Rodrigo}, C., {Solano}, E., \& {Bayo}, A. 2012, {SVO Filter Profile Service
  Version 1.0}, IVOA Working Draft 15 October 2012,
  \dodoi{10.5479/ADS/bib/2012ivoa.rept.1015R}

\bibitem[{{Semenov} \& {Wiebe}(2011)}]{Semenov2011}
{Semenov}, D., \& {Wiebe}, D. 2011, \apjs, 196, 25,
  \dodoi{10.1088/0067-0049/196/2/25}

\bibitem[{{Teague} {et~al.}(2015){Teague}, {Semenov}, {Guilloteau}, {Henning},
  {Dutrey}, {Wakelam}, {Chapillon}, \& {Pietu}}]{Teague2015}
{Teague}, R., {Semenov}, D., {Guilloteau}, S., {et~al.} 2015, \aap, 574, A137,
  \dodoi{10.1051/0004-6361/201425268}

\bibitem[{{Van Der Walt} {et~al.}(2011){Van Der Walt}, {Colbert}, \&
  {Varoquaux}}]{VanDerWalt2011}
{Van Der Walt}, S., {Colbert}, S.~C., \& {Varoquaux}, G. 2011, ArXiv e-prints.
\newblock \doarXiv{1102.1523}

\bibitem[{{{\v{C}}emelji{\'c}} \& {Siwak}(2020)}]{Cemeljic2020MNRAS.491.1057C}
{{\v{C}}emelji{\'c}}, M., \& {Siwak}, M. 2020, \mnras, 491, 1057,
  \dodoi{10.1093/mnras/stz3088}

\bibitem[{{Virtanen} {et~al.}(2019){Virtanen}, {Gommers}, {Oliphant},
  {Haberland}, {Reddy}, {Cournapeau}, {Burovski}, {Peterson}, {Weckesser},
  {Bright}, {van der Walt}, {Brett}, {Wilson}, {Jarrod Millman}, {Mayorov},
  {Nelson}, {Jones}, {Kern}, {Larson}, {Carey}, {Polat}, {Feng}, {Moore}, {Vand
  erPlas}, {Laxalde}, {Perktold}, {Cimrman}, {Henriksen}, {Quintero}, {Harris},
  {Archibald}, {Ribeiro}, {Pedregosa}, {van Mulbregt}, \&
  {Contributors}}]{2019arXiv190710121V}
{Virtanen}, P., {Gommers}, R., {Oliphant}, T.~E., {et~al.} 2019, ArXiv
  e-prints.
\newblock \doarXiv{1907.10121}

\bibitem[{{Woitke} {et~al.}(2019){Woitke}, {Kamp}, {Antonellini}, {Anthonioz},
  {Baldovin-Saveedra}, {Carmona}, {Dionatos}, {Dominik}, {Greaves},
  {G{\"u}del}, {Ilee}, {Liebhardt}, {Menard}, {Min}, {Pinte}, {Rab}, {Rigon},
  {Thi}, {Thureau}, \& {Waters}}]{Woitke2019}
{Woitke}, P., {Kamp}, I., {Antonellini}, S., {et~al.} 2019, \pasp, 131, 064301,
  \dodoi{10.1088/1538-3873/aaf4e5}

\end{thebibliography}
\bibliographystyle{aasjournal}

\appendix

\section{Photospheric modeling with a cold spot}\label{appendix:spotmodel}

To interpret the light curves and constrain the spot position and see a relation between spot position and JVLA continuum emission, we produced a spherical photospheric model.
The \textit{x}-axis was defined along our line-of-sight, while the 
 \textit{y-z} plane is parallel to the projected plane of sky.
The projected photosphere (onto the plane of sky) was resolved in $N^{\it{gird}} \times N^{\it{gird}}$ square cells.
Next, we place a cool spot on the photosphere and rotate it according to the stellar inclination \textit{i}, spot latitude \textit{lat}, longitude \textit{lon}, and rotational phase.
The initial position of the spot is the center of the \textit{y-z} plane.
After we calculate the spot center position, we determined the coordinates of the spot on the sphere that is projected onto the tangent plane at the spot center and whose distance from the spot center is less than or equal to $R_\mathrm{spot}$ in the projection plane. 
The coordinates are used to determine the cells in which the spot is located.
The intensity of each cell is determined by a blackbody model calculated using the temperature of the photosphere and spot.
The linear limb darkening coefficients were taken from \citet{Claret2012}.
The intensities of all grids are added together and the magnitudes are obtained using the information on stellar radius $R_\mathrm{star}$, distance \textit{d}, extinction $A\mbox{v}$, and transmission of the observing filter by \citet{Rodrigo2012} and \citet{Rodrigo2020}.

\begin{figure*}[h]
    \hspace{-1cm}
    \begin{tabular}{ccc}
         \includegraphics[width=6cm]{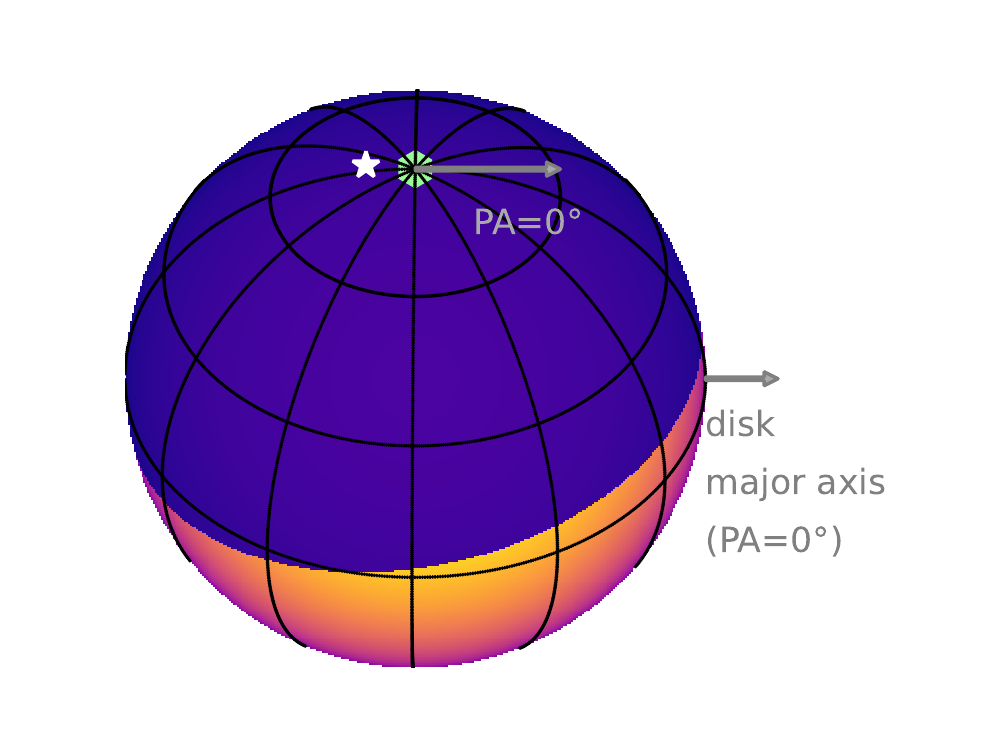} & 
         \includegraphics[width=6cm]{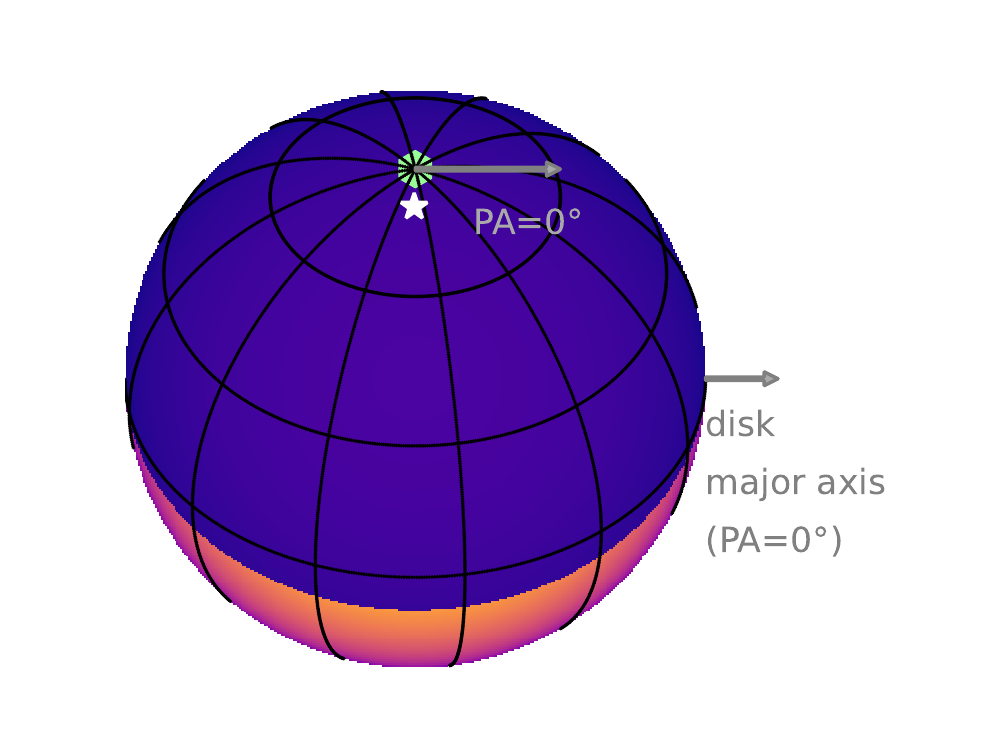} &
          \includegraphics[width=6cm]{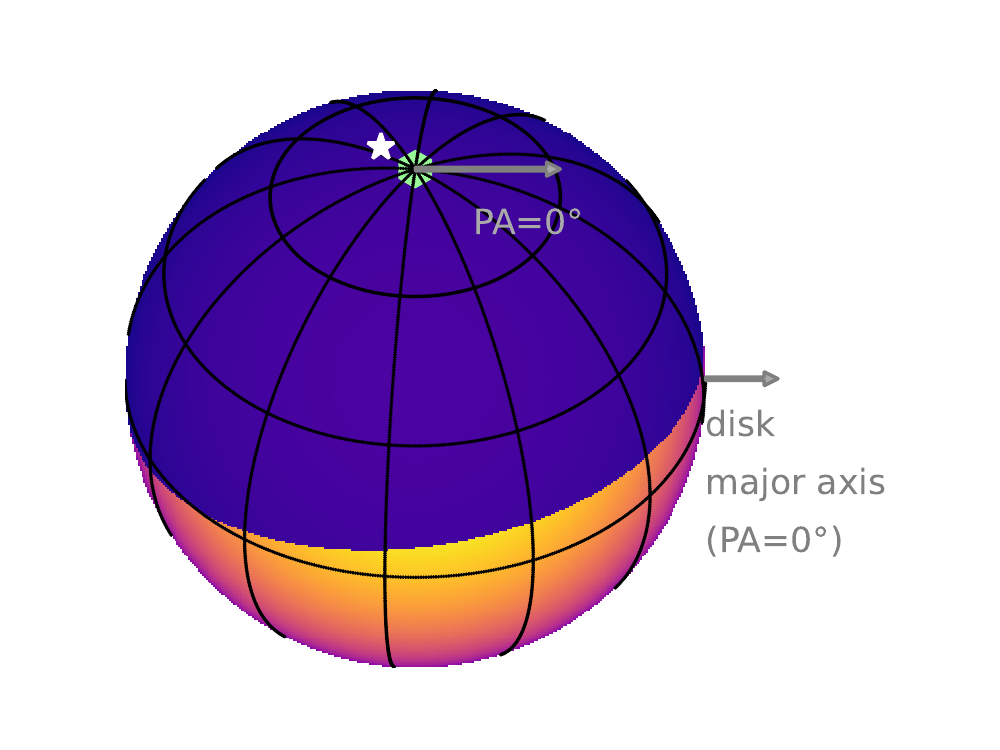} \\ 
    \end{tabular}
    \vspace{-0.5cm}
    \caption{The spot model. The purple region shows a spot. 
    %For simplicity, we used a smaller spot radius than obtained from the fitting results. 
    The orange region shows a photosphere. 
    The color change at the edge is the effect of limb-darkening.
    Grey arrows indicate the direction in which the position angle is zero. 
    The white star indicates the spot center.
    The three vertical lines in Figure \ref{fig:spot_fit} indicate the phase in which each observation was performed. left: Ku-band epoch1, middle: X-band, right: Ku-band epoch3
    }
        \vspace{-0.5cm}
    \label{fig:spotmodel}
\end{figure*}

To pinpoint the free parameters, we performed Markov chain Monte Carlo (MCMC) fittings using the Python package {\tt emcee}.
We fit the model to part of the R and V band photometric data (BJD $<$ 2458527) where periodic modulations were the most obvious. 
Although I-band data is also available, we excluded I-band data from the fit, because low-temperature M-type stars are affected by molecular band emission and atomic emission/absorption around the I-band which complicate the fittings.
In this analysis, we allow the following parameters to be free: $R_\mathrm{star}$, $R_\mathrm{spot}$,  $T_\mathrm{hot}$,  $T_\mathrm{spot}$,\textit{long}, and \textit{lat} as common parameters for all data, where $T_\mathrm{hot}$ and  $T_\mathrm{spot}$ are the non-spotted surface temperatures of the stellar photosphere and the stellar spot, respectively.

Also, we fix stellar rotation period $\mbox{P}_{rot}$, distance \textit{d}, projected rotation velocity $v\sin i$, and the surface gravity $\log g$ and metallicity [Fe/H] for limb-darkening coefficients from the K2 data in Section \ref{sub:period}, the Gaia DR2, \citep{Hartmann1989}, the DIANA project \citep{Woitke2019}, and \citep{Lastennet1999}, respectively. 
We note that the $v\sin i$ is used in combination with $R_\mathrm{star}$ and $\mbox{P}_{rot}$ to determine the stellar inclination\citep{Masuda2020}. 
\begin{equation}
i = \mbox{sin}^{-1}(\frac{v\mbox{sin}i}{v}) = \mbox{sin}^{-1}(\frac{v \mbox{sin} i}{2\pi R_\mathrm{star}\mbox{P}_{rot}}),
\end{equation}

Therefore, the inclination is not included as a free parameter.
In our MCMC fittings, the stellar parameters ($R_\mathrm{star}$ and $T_\mathrm{hot}$) were initialized according to the results reported by the DIANA project \citep{Woitke2019}.
%We imposed a Gaussian prior of $i = 35.2^{+0.7}_{-0.7}$ from \citep{Kudo2018}. 
We initialized $\mbox{T}_{spot}$ to a $\sim$3000 K temperature since \citep{Kospal2018} reported that in M-type protostars, the spot temperature may be a few hundred Kelvin cooler than the temperature of the photosphere. 
We ran the MCMC fittings with 60 walkers and $7 \times 10^3$ steps.
We take the mode of our samples as best fit parameters. 
The uncertainties of the fitting parameters were obtained as the $98\%$ highest posterior density (HPD) interval.
We derived values and uncertainties of the parameters as listed in Table \ref{tab:input_params}.

%TC:endignore
\end{document}